\documentclass[longauth]{aa}
\usepackage{graphicx}
\usepackage{txfonts}
\usepackage[dvipsnames]{xcolor}
\usepackage{orcidlink}
\usepackage{colortbl}

\usepackage{hyperref}  
\hypersetup{colorlinks=true,linkcolor=[rgb]{1.,0.2,0.2},citecolor=[rgb]{0.1,0.4,1.},filecolor=[rgb]{0.7,0.2,0.2},urlcolor=[rgb]{0.7,0.2,0.2}}

\usepackage{color}
\definecolor{blue}{rgb}{0., 0., 1}

\def\Msunyr{\rm{M}_\odot\ \rm{yr}^{-1}}

\def\Msun{M$_\odot$}

\def\Heii{He\,{\sc ii}}
\def\Feii{Fe\,{\sc ii}}
\def\Oiii{[O\,{\sc iii}]}
\def\Civ{C\,{\sc iv}}

\def\Nii{[N\,{\sc ii}]}

\def\Cii{[C\,{\sc ii}]}

\def\Ha{H$\alpha$}
\def\Hb{H$\beta$}
\def\Hg{H$\gamma$}
\def\Hd{H$\delta$}
\def\Mgii{Mg\,{\sc ii}}
\def\Civ{C\,{\sc iv}}
\def\Sii{[S\,{\sc ii}]}
\def\Siiv{Si\,{\sc iv}}
\def\kms{km\,s$^{-1}$}

\def\lsim{\mathrel{\rlap{\lower 3pt \hbox{$\sim$}} \raise 2.0pt \hbox{$<$}}}
\def\gsim{\mathrel{\rlap{\lower 3pt \hbox{$\sim$}} \raise 2.0pt \hbox{$>$}}}

\begin{document}

\authorrunning{Loiacono et al.}
\titlerunning{Black hole mass from the NIRSpec spectrum of a $z=6.2$ quasar}
\title{A quasar-galaxy merger at $z\sim6.2$: black hole mass and quasar properties from the NIRSpec spectrum}
\author{
Federica Loiacono\inst{1}$^{\orcidlink{0000-0002-8858-6784}}$, Roberto Decarli\inst{1}$^{\orcidlink{0000-0002-2662-8803}}$, Marco Mignoli\inst{1}$^{\orcidlink{0000-0002-9087-2835}}$, Emanuele Paolo Farina\inst{2}$^{\orcidlink{0000-0002-6822-2254}}$, Eduardo Ba$\tilde{\rm n}$ados\inst{3}$^{\orcidlink{0000-0002-2931-7824}}$, Sarah Bosman\inst{4, 3}$^{\orcidlink{0000-0001-8582-7012}}$, Anna-Christina Eilers\inst{5}$^{\orcidlink{0000-0003-2895-6218}}$, Jan-Torge Schindler\inst{6}$^{\orcidlink{0000-0002-4544-8242}}$, Michael A. Strauss\inst{7}$^{\orcidlink{0000-0002-0106-7755}}$, Marianne Vestergaard\inst{8, 9}$^{\orcidlink{0000-0001-9191-9837}}$, Feige Wang\inst{9}$^{\orcidlink{0000-0002-7633-431X}}$, Laura Blecha\inst{10}$^{\orcidlink{0000-0002-2183-1087}}$, Chris L. Carilli\inst{11}$^{\orcidlink{0000-0001-6647-3861}}$, Andrea Comastri\inst{1}$^{\orcidlink{0000-0003-3451-9970}}$, Thomas Connor\inst{12, 13}$^{\orcidlink{0000-0002-7898-7664}}$, Tiago Costa\inst{14}$^{\orcidlink{0000-0002-6748-2900}}$, Massimo Dotti\inst{15, 16, 17}$^{\orcidlink{0000-0002-1683-5198}}$, Xiaohui Fan\inst{9}$^{\orcidlink{0000-0003-3310-0131}}$, Roberto Gilli\inst{1}$^{\orcidlink{0000-0001-8121-6177}}$, Hyunsung D. Jun\inst{18}$^{\orcidlink{0000-0003-1470-5901}}$, Weizhe Liu\inst{9}$^{\orcidlink{0000-0003-3762-7344}}$, Alessandro Lupi\inst{15}$^{\orcidlink{0000-0001-6106-7821}}$, Madeline A. Marshall\inst{19, 20}$^{\orcidlink{0000-0001-6434-7845}}$, Chiara Mazzucchelli\inst{21}$^{\orcidlink{0000-0002-5941-5214}}$, Romain A. Meyer\inst{22}$^{\orcidlink{0000-0001-5492-4522}}$, Marcel Neeleman\inst{23}$^{\orcidlink{0000-0002-9838-8191}}$, Roderik Overzier\inst{24}$^{\orcidlink{0000-0002-8214-7617}}$, Antonio Pensabene\inst{15}$^{\orcidlink{0000-0001-9815-4953}}$, Dominik A. Riechers\inst{25}$^{\orcidlink{0000-0001-9585-1462}}$, Benny Trakhtenbrot\inst{26}$^{\orcidlink{0000-0002-3683-7297}}$, Maxime Trebitsch\inst{27}$^{\orcidlink{0000-0002-6849-5375}}$, Bram Venemans\inst{24}$^{\orcidlink{0000-0001-9024-8322}}$, Fabian Walter\inst{3}$^{\orcidlink{0000-0003-4793-7880}}$, Jinyi Yang\inst{9}$^{\orcidlink{0000-0001-5287-4242}}$
}
\institute{
INAF -- Osservatorio di Astrofisica e Scienza dello Spazio di Bologna, via Gobetti 93/3, I-40129, Bologna, Italy. \email{ federica.loiacono1@inaf.it} \and Gemini Observatory, NSF’s NOIRLab, 670 N A’ohoku Place, Hilo, Hawai’i 96720, USA \and
Max Planck Institut f\"{u}r Astronomie, K\"{o}nigstuhl 17, D-69117, Heidelberg, Germany \and Institute for Theoretical Physics, University of Heidelberg, Philosophenweg 16, 69120 Heidelberg, Germany \and MIT Kavli Institute for Astrophysics and Space Research, 77 Massachusetts Ave., Cambridge, MA 02139, USA \and Hamburger Sternwarte, Universität Hamburg, Gojenbergsweg 112, D-21029 Hamburg, Germany \and Department of Astrophysical Sciences, Princeton University, Princeton, NJ 08544 USA \and The Niels Bohr Institute, University of Copenhagen, Denmark \and Steward Observatory, University of Arizona, 933 N Cherry Avenue, Tucson, AZ 85721, US \and Department of Physics, University of Florida, Gainesville, FL 32611-8440, USA \and National Radio Astronomy Observatory, P.O. Box O, Socorro, NM 87801, USA \and Center for Astrophysics | Harvard \& Smithsonian, 60 Garden St., Cambridge, MA 02138, USA \and Jet Propulsion Laboratory, California Institute of Technology, 4800 Oak Grove Drive, Pasadena, CA 91109, USA \and Max-Planck-Institut f\"{u}r Astrophysik, Karl-Schwarzschild-Straße 1, D-85748 Garching b. M\"{u}nchen, Germany \and Dipartimento di Fisica “G. Occhialini”, Università degli Studi di Milano-Bicocca, Piazza della Scienza 3, I-20126 Milano, Italy \and INFN, Sezione di Milano-Bicocca, Piazza della Scienza 3, I-20126 Milano, Italy \and INAF - Osservatorio Astronomico di Brera, via Brera 20, I-20121 Milano, Italy \and Department of Physics, Northwestern College, 101 7th St SW, Orange City, IA 51041 \and National Research Council of Canada, Herzberg Astronomy \& Astrophysics Research Centre, 5071 West Saanich Road, Victoria, BC V9E 2E7, Canada \and ARC Centre of Excellence for All Sky Astrophysics in 3 Dimensions (ASTRO 3D), Australia \and Instituto de Estudios Astrof\'isicos, Facultad de Ingenier\'ia y Ciencias, Universidad Diego Portales, Avenida Ejercito Libertador 441, Santiago, Chile \and Department of Astronomy, University of Geneva, Chemin Pegasi 51, 1290 Versoix, Switzerland \and National Radio Astronomy Observatory, 520 Edgemont Road, Charlottesville, VA, 22903, USA \and Leiden Observatory, Leiden University, Niels Bohrweg 2, NL-2333 CA Leiden, Netherlands \and I. Physikalisches Institut, Universität zu Köln, Zülpicher Strasse 77, 50937 Köln, Germany \and School of Physics and Astronomy, Tel Aviv University, Tel Aviv 69978, Israel \and Kapteyn Astronomical Institute, University of Groningen, P.O. Box 800, 9700 AV Groningen, The Netherlands
}
\date{}
\abstract{We present JWST/NIRSpec integral field data of the quasar PJ308-21 at $z = 6.2342$. As shown by previous ALMA and HST imaging, the quasar has two companion sources, interacting with the quasar host galaxy. The high-resolution G395H/290LP NIRSpec spectrum covers the $2.87 - 5.27\ \rm \mu m$ wavelength range and shows the rest-frame optical emission of the quasar with exquisite quality ($S/N \sim 100 - 400$ per spectral element). Based on the \Hb\ line from the broad line region, we obtain an estimate of the black hole mass $M_{\rm BH, H\beta} \sim 2.7 \times 10^{9}\ \rm M_{\odot}$. This value is within a factor $\lesssim 1.5$ of the \Ha -based black hole mass from the same spectrum ($M_{\rm BH, H\alpha} \sim 1.93 \times 10^{9}\ \rm M_{\odot}$) and is consistent with a previous estimate relying on the \Mgii\ $\lambda 2799$ line ($M_{\rm BH, MgII} \sim 2.65 \times 10^{9}\ \rm M_{\odot}$). All these $M_{\rm BH}$ estimates are within the $\sim 0.5$ dex intrinsic scatter of the adopted mass calibrations. 
The high Eddington ratio of PJ308-21 $\lambda_{\rm Edd, H\beta} \sim 0.67$ ($\lambda_{\rm Edd, H\alpha} \sim 0.96$) is in line with the overall quasar population at $z \gtrsim 6$.
The relative strengths of the \Oiii , \Feii\ and \Hb\ lines are consistent with the empirical "Eigenvector 1" correlations as observed for low redshift quasars.
We find evidence for blueshifted \Oiii\ $\lambda 5007$ emission with a velocity offset $\Delta v_{\rm [O\ III]} = -1922\pm 39$ \kms\ from the systemic velocity and a $\rm FWHM($\Oiii$) = 2776^{+75}_{-74}$ \kms . This may be the signature of outflowing gas from the nuclear region, despite the true values of $\Delta v_{\rm [O\ III]}$ and $\rm FWHM($\Oiii$)$ are likely more uncertain due to the blending with \Hb\ and \Feii\ lines. Our study demonstrates the unique capabilities of NIRSpec in capturing quasar spectra at cosmic dawn and studying their properties in unprecedented detail.}    
\maketitle    

\section{Introduction} 
High-redshift quasars (or quasi-stellar objects, QSOs) are key laboratories for studying the early stages of galaxy evolution and black hole formation (see \citealt{fan23} for a recent review).
At $z \gtrsim 6$, i.e., when the Universe was $\lesssim 1\rm\ Gyr$ old, they are often harbored by galaxies with star formation rates $\rm{SFR} \sim 100 - 4000\ \Msunyr $, significantly higher than typical galaxies at similar redshifts with no nuclear activity (e.g., \citealt{vanzella14, bouwens15, salmon15, ota17, harikane18, walter22}). 
The central black holes can already have masses of $10^{9}$ \Msun\ (e.g., \citealt{yang21, farina22, mazzucchelli23}) and the molecular gas content of the host galaxy, which feeds both nuclear and star formation activity, can easily exceed $10^{10}$\ \Msun\ (e.g., \citealt{walter03, wang08, venemans17}). 
Accounting for the rapid growth of $z \gtrsim 6$ quasars poses important challenges to models of early black hole formation and evolution (see, e.g., the reviews by \citealt{inayoshi20} and \citealt{volonteri21}).\\
\indent Over the last years, significant effort has been invested to study $z \gtrsim 6$ QSOs using the most advanced facilities. In particular, observations with the \textit{Atacama Large Millimeter/submillimeter Array} (ALMA) and the \textit{Northern Extended Millimeter Array} (NOEMA) have provided a detailed view of the rest-frame far-infrared emission in these sources. Sub-mm and mm data have enabled deep insights into several properties, such as the gas and dust content \citep{wang08, calura14, decarli18, venemans18, decarli22}, kinematics (e.g., \citealt{willott13, willott15, venemans16, neeleman19, pensabene20, neeleman21}) and the physics of the interstellar medium (ISM; \citealt{walter18, novak19, li20, pensabene21, meyer22, shao22, tripodi22, decarli23}) of quasar host galaxies.
Observations at these wavelengths have revealed that $z \gtrsim 6$ QSOs are often associated with companion galaxies from small to large distances, with overdensities exceeding by several orders of magnitude the expectations for the field population of galaxies (\citealt{aravena16, decarli17, meyer22over}; see also \citealt{zana23} for the theoretical interpretation), even if there are also exceptions to this trend \citep{mazzucchelli17, habouzit19}.\\
\indent Observations with ground-based optical/near-infrared facilities have constrained the rest-frame ultraviolet (UV) emission from $z \gtrsim 6$ quasars (e.g., \citealt{kurk07, willott10, mazzucchelli17a, onoue19, schindler20, yang21}). For example, integral field data from the Multi Unit Spectroscopic Explorer (MUSE) show that quasars are often surrounded by Lyman $\alpha$ nebulae extending up to $10-30\ \rm kpc$ (e.g., \citealt{borisova16, farina17, farina19, drake19}). Using X-Shooter detections of the \Mgii\ $\lambda 2799$ and \Civ\ $\lambda 1549$ broad emission lines from quasars, \citet{farina22} provided estimates of black hole masses (see also \citealt{derosa14, mazzucchelli23}), which enabled the study of the scaling relations between black hole and galaxy mass at $z \gtrsim 6$ (e.g., \citealt{willott17, izumi19}; see also \citealt{pensabene20, neeleman21}).\\
\indent Despite the significant effort with the most advanced ground telescopes that has been invested over the last years, a number of questions remains unsolved.
In particular, the rest-frame optical lines from $z > 6$ quasars, shifted to the near- and mid-infrared at those redshifts, represent an uncharted territory due to the limits of the observing facilities in that window.\\
\indent Now, with the \textit{James Webb Space Telescope} (JWST; \citealt{gardner23}) it is finally possible to study the rest-frame optical emission from $z > 6$ quasars. Observations of emission lines originating from the broad line region (BLR; e.g., \Ha\ and \Hb ) provide robust estimates of the black hole mass based on local relations \citep{2005ApJ...630..122G, vestpet06, shen11}.
In addition, the strength and ratios of nebular lines such as \Ha , \Hb , \Oiii\ $\lambda \lambda 4959, 5007$, \Nii\ $\lambda \lambda 6548, 6584$ offer a unique tool to characterize the ISM of the quasar host and companion galaxies, constraining important physical quantities (e.g., the ionization mechanism and the ionization parameter, metallicity, dust extinction), which are not accessible at these redshifts using UV spectra only.   
First promising results on some of the $z > 6$ quasars observed with JWST come from the "A SPectroscopic survey of biased halos In the Reionization Era" project (ASPIRE; \citealt{wang23}). Based on the ASPIRE data, \citet{yang23} studied the rest-frame emission at $4100-5100$ \AA\ in a sample of eight $z > 6.5$ quasars while \citet{wang23} discovered a filamentary structure tracing an overdensity around a luminous $z \sim 6.6$ QSO (see also \citealt{kashino23}). \citet{marshall23} studied two other $z \sim 6.8$ quasars, as well as their host galaxies and companion sources. These works also provide first estimates of black hole masses at $z > 6$ measured with the \Hb\ or \Ha\ line (see also \citealt{bosman23, ding23, eilers23, yue23}).\\
\indent PJ308-21 at $z = 6.2342$ stands out among the known quasars at $z > 6$ as a key laboratory to study quasar-galaxy interactions at cosmic dawn. This object was discovered by \citet{banados16} (see also \citealt{mazzucchelli17a}) and intensively studied by \citet{decarli19} using high-resolution ALMA data, targeting the \Cii\ and dust continuum emission in this system. The ALMA observations reveal that the quasar host has two companion sources, respectively blueshifted ($\Delta v_{\rm [CII]} \sim - 750$ \kms ; \citealt{decarli19}) and redshifted ($\Delta v_{\rm [CII]} \sim + 500$ \kms) from the quasar, located toward the western and eastern sides of the quasar at projected distances of $\sim 5$ and $\gtrsim 10\ \rm kpc$ respectively. These two emission components are also detected in rest-frame UV images from the \textit{Hubble Space Telescope} (HST), revealing light from young stars from both objects.
As discussed by \citet{decarli19}, the quasar host is possibly experiencing a merger with another galaxy, tidally disturbed by the interaction and which accounts for the emission from the two companion sources.
Deep MUSE observations reveal that the system is surrounded by a Lyman $\alpha$ halo ($R \sim 27$ kpc), which overlaps with the eastern companion \citep{farina19}. Furthermore, \citet{connor19} found that PJ308-21 has an X-rays luminosity $L_X(\rm 2-10\ keV) \sim 2 \times 10^{45}\ \rm erg\ s^{-1}$ while the western companion shows possible high-energy emission with no soft component, which may point to an AGN nature also for this object. 
Because of its complex and extended morphology, the PJ308-21 system is an ideal target for JWST/NIRSpec integral field spectroscopy, which allow us to probe rest-frame optical emission of the spatially resolved components of this object.\\
\indent This work (Paper I) is the first of a series of papers studying the PJ308-21 system with NIRSpec data. In particular, it is devoted to the characterization of the quasar spectrum itself. Paper II (Decarli et al., in prep.) studies the extended gas emission in the system. Paper III (Farina et al. in prep.) investigates the Lyman $\alpha$/\Ha\ halo. Yet another paper will focus on the kinematics of the system.\\
\indent This paper is organized as follows. Sect.~\ref{sec:datared} deals with the NIRSpec observation and data reduction. Sect.~\ref{SEC:analysis} is devoted to the analysis of the QSO spectrum. 
In Sect.~\ref{sec:results} we present the results of our analysis. Finally, we summarize our conclusions in Sect.~\ref{sec:concl}.\\
\indent In this work we adopt a $\Lambda$CDM cosmology with $\Omega_{\Lambda} = 0.7$, $\Omega_{\rm M} = 0.3$ and a Hubble constant $H_0 = 70\ \rm km\ s^{-1}\ Mpc^{-1}$.

\section{Observations and data reduction}
\label{sec:datared}
\begin{figure*}
\begin{center}
\includegraphics[width=0.75\textwidth]{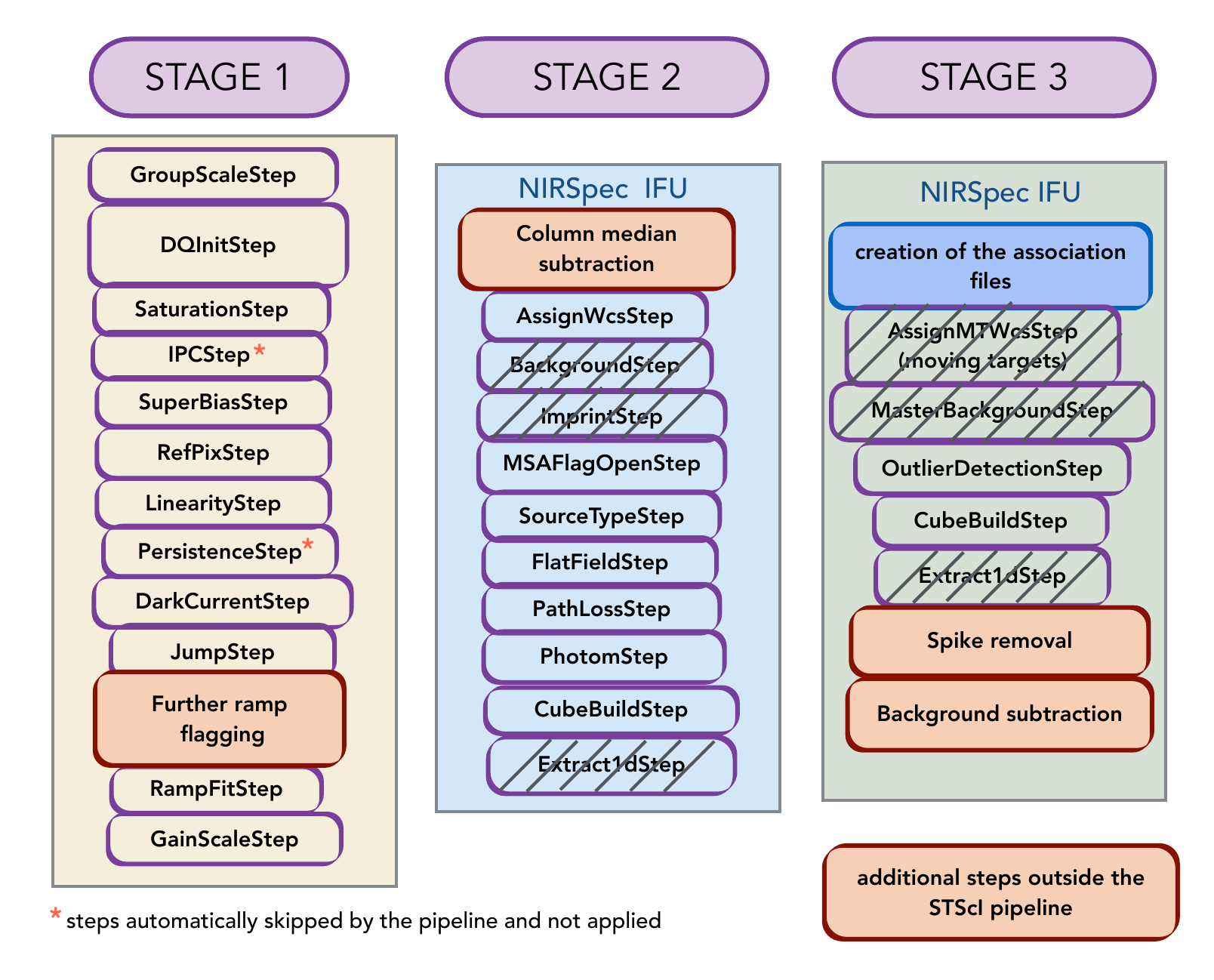}\\
\end{center}
\caption{Scheme of the data reduction process for the NIRSpec data of PJ308-21. The data reduction is divided into three stages, each of them consisting of several steps. Stage 1 is independent from the JWST instrument and mode, unlike stage 2 and 3, for which we present the steps for NIRSpec/IFU.
We skip some pipeline steps (boxes with grey tilted lines) and introduce additional steps (orange boxes) to improve the data reduction (see Section~\ref{sec:datared}).}
\label{fig_red}
\end{figure*}
In this paper we analyze the data of the Cycle 1 General Observers (GO) program 1554 (PI: R. Decarli) that targets the PJ308-21 quasar system (see Table 1 of \citealt{decarli19} for the properties of the system derived from ALMA and HST data). The observations were taken on September 22-23 2022 using the integral field unit (IFU) of JWST/NIRSpec \citep{jako22, boker22}. The science observations consist of two separated fields ($3\arcsec \times 3\arcsec$ each), corresponding to the quasar and its western companion (PJ308-W), and to the eastern component (PJ308-E) respectively. A third field covered an empty region of the sky to estimate the background. We observed each field (background included) for 2 hours and 30 minutes (including overheads), for a total execution time of 7.5 hours (4.8 hrs on source). We used the NRSIRS2 readout in order to minimize the correlated noise and the data volume, and observed a total of 18 exposures (6 per pointing applying a small-cycling dithering pattern), each of them consisting of 13 groups (i.e., integration elements). We also observed one leakage exposure of $87$ seconds to remove the pattern introduced by the NIRSpec Micro-Shutter Array (MSA).
 \\
\indent The data were taken in the high-resolution mode (spectral power $R\sim 2700$) using the G395H/290LP grating-filter combination. This covers the wavelength range $2.87 - 5.27\ \rm \mu m$ with a gap due to the NRS1 and NRS2 detector separation corresponding to $4.1 - 4.3\ \rm \mu m$ .
\\
\indent We reduced the data using the JWST pipeline developed by the Space Telescope Science Institute (STScI). We use the 1.12.5 version of the pipeline combined with the Calibration Reference Data System (CRDS) context \texttt{jwst\_1183.pmap}, which indicates the set of reference files necessary for calibrating the data.
The reduction process consists of three stages, each of them involving several steps (see Figure~\ref{fig_red}). Stage 1 processes the raw exposures (i.e., non-destructive readout ramps) correcting for detector effects like bias, dark and cosmic rays impact, and creates count rate images. 
During this stage, we introduce a further flagging of the ramps strongly affected by saturation, non-linearity or cosmic rays (\texttt{Further ramp flagging} in Figure~\ref{fig_red}). In particular, we exclude from the ramp fitting all the ramps containing only two useful groups (out of 13). This step results in a drop of the standard deviation in the rate files (i.e., 2D count-rate images) by a factor $\lesssim 2 \times 10^3$. 
\begin{figure*}
\begin{center}
\includegraphics[width=0.7\textwidth]{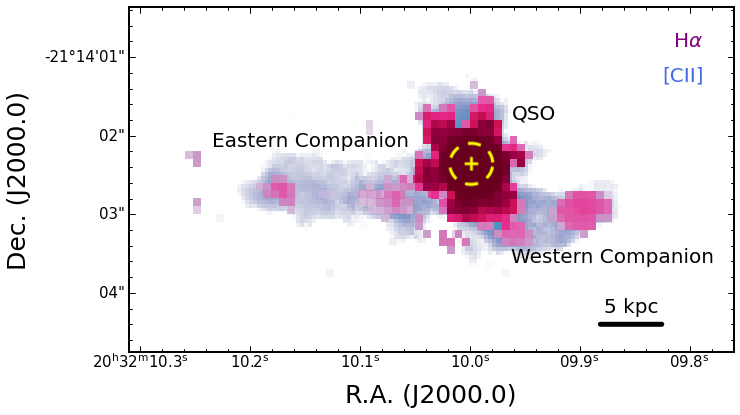}\\
\end{center}
\caption{The PJ308-21 system. The quasar has two companion sources, located to the western and eastern sides of the QSO respectively. We show the \Ha\ and the underlying continuum emission from the NIRSpec/IFU data (magenta) overimposed on the \Cii\ emission from ALMA high-resolution data (blue; \citealt{decarli19}). The yellow cross and circle show the aperture for the spectrum extraction (see Sect.~\ref{sub:apcor}).}
\label{fig_system}
\end{figure*}
The rate files are affected by vertical stripes due to residual $1/f$ noise unremoved in stage 1 (see also \citealt{kashino23, perna23, rauscher23}). We eliminate this pattern by subtracting from each spectral column in the 2D count-rate images its median value, which provides a reasonable estimate of any spectrally-dependent correlated noise (see \texttt{Column median subtraction} in Figure~\ref{fig_red}).
\indent We thus process the median-subtracted exposures through stage 2 of the reduction, where flat field correction, wavelength and flux calibration are applied to all the exposures. We skip the \texttt{BackgroundStep}, as we apply the background subtraction at the end of stage 3.
We also skip the \texttt{ImprintStep}, which is meant to subtract the pattern created by the MSA, as we found that the subtraction of a dedicated leakage exposure results in a detrimental increase of the noise in the rate files. 
\\
\indent We then perform stage 3, where all the exposures are combined into the final datacubes. We apply the \texttt{OutlierDetectionStep}, which is meant to remove the remaining bad pixels from the calibrated exposures, and which provides reliable results for our dataset from the 1.11.4 pipeline version.\\ 
\indent The one-dimensional spectra extracted from random apertures in the final datacubes show several spikes, due to cosmic rays and possible wrong solutions for the ramp fitting.
We filter out these spikes by masking all the voxels that present a jump between contiguous channels persisting for less than four channels (corresponding to $< 150$ \kms\ at $4.0\ \mu \rm m$), which is the typical width of the spikes. The procedure is applied to all the voxels of the datacubes (see \texttt{Spike removal} in Figure~\ref{fig_red}). We carefully inspect the voxels corresponding to the science targets before/after applying this step to check that none of the emission lines is affected by the algorithm.
Finally, we subtract the background, which is higher for the spectral region sampled by the NRS2 detector ($4.3 - 5.3 \ \rm \mu m$), as the NIR background grows with increasing wavelength \citep{rigby22back}. 
We estimate the background from the median of spectra extracted from empty regions in the FOV as a function of wavelength.
The median background is then subtracted from the datacubes channel-by-channel (see \texttt{Background subtraction} in Figure~\ref{fig_red}).\\ 
\indent In this paper, as we are interested in the quasar spectrum, we study the PJ308-W pointing only. A map showing an overlook on the system captured by the PJ308-W and PJ308-E reduced datacubes is shown in Figure~\ref{fig_system} (see Decarli et al., in prep., for more details).
\section{Analysis}
\label{SEC:analysis}
\subsection{Spectrum extraction and aperture correction}
\label{sub:apcor}
\begin{figure*}
\begin{center}
\includegraphics[width=0.9\textwidth]{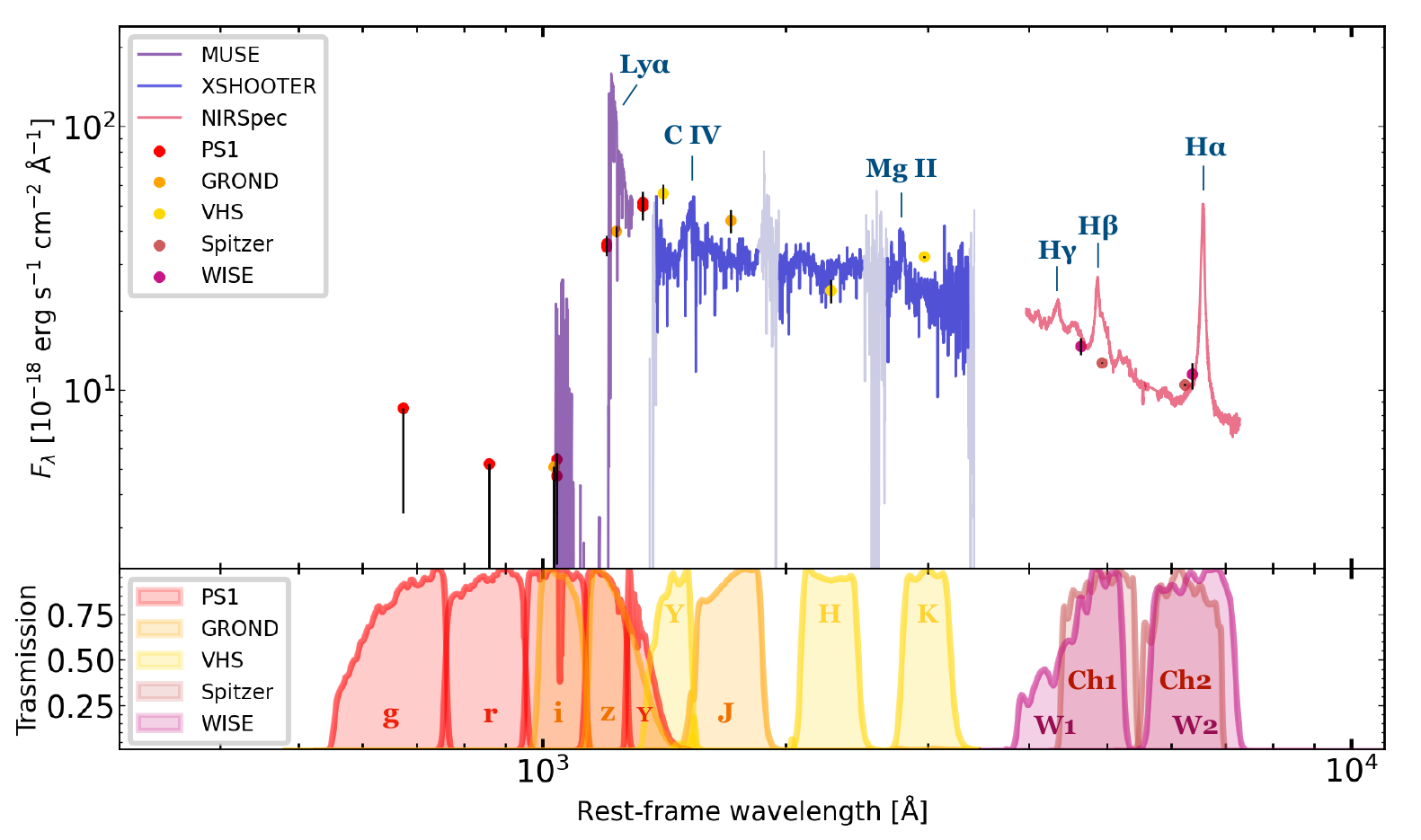}\\
\end{center}
\caption{Rest-frame UV/optical spectroscopic and photometric data of PJ308-21. The main emission lines are indicated. The aperture-corrected NIRSpec spectrum is shown in red. The purple and blue lines show MUSE and XSHOOTER spectra from previous works \citep{farina19, schindler20}. We show in grey the XSHOOTER data affected by the atmosphere absorption. The photometric data from Pan-STARRS1 (PS1; \textit{g, r, i, z, Y} bands), GROND (\textit{i, z, J} bands), VHS (\textit{Y, H, K} bands), \textit{Spitzer} (\textit{Ch1}=3.5 $\rm \mu m$, \textit{Ch2}=4.5 $\rm \mu m$) and WISE (\textit{W1}=3.3 $\rm \mu m$, \textit{W2}=4.6 $\rm \mu m$) are shown with the associated errors \citep{banados16}.}
\label{fig_spec}
\end{figure*}

We compute a wavelength-dependent aperture correction based on the observation of a point source (i.e., a star) using archival data.
We use the uncalibrated exposures of the GSPC P330-E G--dwarf, observed in the 1538 calibration program (PI: K. Gordon). We opt for this star as it was observed with NIRSpec/IFU using the same grating/filter combination of our observation. We process the data, applying the same steps described in Sect.~\ref{sec:datared}. The final datacube is used for extracting one-dimensional spectra adopting circular apertures with radii $R$ ranging from $R = 0.025\arcsec$ to $R = 1.2\arcsec$ centered on the star. 
The growth curve asympotes at $1 \arcsec$, so we normalize the flux in each wavelength channel to the value at $R = 1\arcsec$ (see also Fig~\ref{fig_psfProf} in Appendix~\ref{app:aper}).
We thus estimate the flux fraction for fixed aperture radii as a function of the wavelength (see Table~\ref{tab:apcorr} for representative wavelengths over the G395H/F290LP spectral range).
We compute these values by fitting a second--order spline to the flux fractions. The enclosed flux fraction increases with increasing radius and, for fixed aperture, decreases with wavelength. The latter trend is less pronounced at larger radii, as most of the flux is also included for the redder part of the spectrum (see also Fig~\ref{fig_psfProf} in Appendix~\ref{app:aper}). 
For an aperture of $0.3\arcsec$, the flux loss between the red and blue channels can be as high as $\sim 6 \%$, while the difference is lower for larger apertures.\\  
\indent The values reported in Table~\ref{tab:apcorr} can be used to correct the spectra extracted from apertures with radii $R \sim 0.3\arcsec - 0.7\arcsec$ to recover the total flux emitted by a point source. These values should provide reliable corrections when using the dataset observed using the NIRSpec/IFU with the G395H/F290LP grating/filter combination and the adopted version of the pipeline\footnote{We also test the procedure on another unresolved source, IRAS 05248-7007, observed with the G395H/F290LP grating/filter setup in the NIRSpec calibration program 1492 (PI: T. Beck). For apertures of $0.3"-0.7"$ we obtain flux fractions that differ from the values reported in Table~\ref{tab:apcorr} by less than $1 \%$.}.
We adopt an extraction aperture of $0.3\arcsec$ centered at the NIRSpec peak position of PJ308-21 (Figure~\ref{fig_system}). We note that using a larger aperture, i.e., $0.4\arcsec - 0.7\arcsec$, after applying the aperture correction gives perfectly consistent results, while using a smaller aperture ($\sim 0.2\arcsec$) cannot properly recover the QSO flux because of the undersampling of NIRSpec PSF (see \citealt{boker22}). We thus apply the 0.3\arcsec\ aperture correction based on Table~\ref{tab:apcorr} to recover the total flux. The same correction is applied to the error spectrum. We obtain the noise spectrum by summing all the pixel variance over the 0.3\arcsec\ aperture and extracting the square root of the summed total. 
We  show in Fig.~\ref{fig_spec} the aperture-corrected spectrum together with the rest-frame UV spectra of PJ308-21 from the literature, obtained with MUSE (\citealt{farina19}) and XSHOOTER \citep{schindler20}. The $S/N$ of the NIRSpec spectrum per spectral element is $100 - 400$. 
We show also the photometric data (see \citealt{banados16}) from the first release of the Panoramic Survey Telescope and Rapid Response System (Pan-STARRS1, PS1; \textit{g, r, i, z, Y} bands), the Gamma-ray Burst Optical/Near-infrared Detector (GROND; \textit{i, z, J} bands), the VISTA Hemisphere Survey (VHS; \textit{Y, H, K} bands), \textit{Spitzer (\textit{Ch1}=3.5 $\rm \mu m$, \textit{Ch2}=4.5 $\rm \mu m$)} and the Wide-Field Infrared Survey Explorer (WISE; \textit{W1}=3.3 $\rm \mu m$, \textit{W2}=4.6 $\rm \mu m$). There is an overall agreement between the NIRSpec spectrum and the available photometry from \textit{Spitzer} and WISE. The differences, in the order of $\sim10\%$, are possibly due to systematics in the instruments zero-points.
We note that the continuum slope changes between the NIRSpec data and the ground-based spectra. This is possibly due to both physical and instrumental effects. The power-law continuum in quasars is interpreted as the result of several black bodies, corresponding to layers in the accretion disk with different temperatures. The large black hole mass (see Sect.~\ref{sub:bhmass}) yields a relatively large radius of the innermost ring, which corresponds to a peak in the continuum emission. Therefore, the quasar continuum is expected to change the slope from the rest-frame UV (where most of the disk emission is expected) to rest-frame optical. Second, the fluxes in the XSHOOTER spectrum are expected to be more extinguished from the dust in the quasar host in comparison to the near-infrared data, which can also influence the slope. Beyond these physical effects, possible systematics in the zero-points, potential slit losses in the XSHOOTER spectrum, uncertainties in the NIRSpec flux calibration and increased noise toward the red end of the XSHOOTER spectrum can also contribute to the slope variation.

\subsection{Spectral fitting}
\label{sub:fit}
We convert the observed wavelength into the rest-frame using the \Cii\ based redshift ($z = 6.2342$, \citealt{decarli19}). The spectrum shows a strong quasar continuum and all the prominent Balmer lines (H$\alpha$, H$\beta$, H$\gamma$, H$\delta$) due to the BLR. We also observe bright emission from the \Feii\ multiplets in the $4000 - 5500\ \AA$ range. The BLR dominates over the nebular emission of the narrow line region (NLR; i.e., \Oiii\ $\lambda \lambda 4959, 5007$ and \Nii\ $\lambda \lambda 6548, 6584$ doublets). We do not detect a significant \Heii\ $\lambda 4686$ emission line, which can be used as a tracer of the nuclear activity \citep{2023MNRAS.521.1264T}, or significant \Sii\ $\lambda \lambda 6716, 6732$ emission. Also, we do not find absorption features over the NIRSpec spectrum.
\begin{table}
\caption{Flux fraction for fixed aperture radius $R$ as a function of the wavelength (NIRSpec/IFU, G395H/F290LP).}              
\label{tab:apcorr}     
\centering                                      
\scalebox{0.9}{
\begin{tabular}{c c c c c c}          
\hline\hline       

Wavelength & 0.3\arcsec & 0.4\arcsec & 0.5\arcsec & 0.6\arcsec & 0.7\arcsec \\    
($\mu$m)  & & & & & \\
\hline                                   
    2.90 & 0.891 & 0.926 & 0.953 & 0.964 & 0.973 \\
    3.00 & 0.891 & 0.926 & 0.952 & 0.964 & 0.974 \\
    3.20 & 0.890 & 0.925 & 0.950 & 0.965 & 0.976 \\
    3.40 & 0.888 & 0.924 & 0.948 & 0.966 & 0.977 \\
    3.60 & 0.885 & 0.923 & 0.947 & 0.966 & 0.977 \\
    3.80 & 0.882 & 0.922 & 0.945 & 0.966 & 0.978 \\
    4.00 & 0.877 & 0.921 & 0.944 & 0.965 & 0.978 \\
    4.20 & 0.872 & 0.920 & 0.942 & 0.963 & 0.978 \\
    4.40 & 0.866 & 0.919 & 0.941 & 0.962 & 0.977 \\
    4.60 & 0.859 & 0.918 & 0.940 & 0.959 & 0.976 \\
    4.80 & 0.852 & 0.916 & 0.938 & 0.957 & 0.975 \\
    5.00 & 0.843 & 0.915 & 0.937 & 0.954 & 0.973 \\
    5.10 & 0.839 & 0.914 & 0.936 & 0.952 & 0.973 \\
    5.20 & 0.834 & 0.914 & 0.936 & 0.950 & 0.972 \\
\hline                                             
\end{tabular}
}
\end{table}
\subsubsection{Continuum subtraction}
\label{subsub:cont}
We perform a multicomponent fitting to reproduce the NIRSpec spectrum using a Markov chain Monte Carlo (MCMC) method with the Python package \texttt{emcee} \citep{2013PASP..125..306F}. We first model the continuum emission, whereafter we model the iron and the nebular lines.\\
\indent Quasar spectra often show a break in their shape, possibly due to the host galaxy contamination (see, for example, \citealt{vb01}). For this reason, we use a broken power-law to model the continuum emission, in the functional form of 
\small
\begin{equation*}
f(\lambda) = 
\begin{cases}
A \left( \dfrac{\lambda}{\lambda_{\rm break} } \right)^{-\alpha_1} : \lambda < \lambda_{\rm break} \\
A \left( \dfrac{\lambda}{\lambda_{\rm break}} \right)^{-\alpha_2} : \lambda \geq \lambda_{\rm break}
\end{cases}
\end{equation*}
\normalsize
where $A$ is the normalization, $\lambda_{\rm break}$ is the break wavelength and $\alpha_1$ and $\alpha_2$ are the power-law indices. 
Following \citet{vb01}, we use the line-free regions of the spectra $6005-6035\ \AA$ and $7160-7180\ \AA$ with an additional region at $4000-4030\ \AA$ to constrain the bluest part of the spectrum, after carefully checking that it is not contaminated by line emission (see the shaded regions in Figure~\ref{fig_fit_cont}, top panel) to fit the continuum.
The continuum model is shown in Figure~\ref{fig_fit_cont} (top panel), with the continuum-subtracted spectrum highlighted in orange (see also Table~\ref{tab:fitcontiron} for the fitted parameters). We do not discuss the physical meaning of the $\lambda_{\rm break}$, $A$, $\alpha_1$ and $\alpha_2$ values and compare with low-redshift quasars, as we find that the parameters are highly degenerate among themselves. We thus use the fit only to obtain a robust continuum-free spectrum for fitting the emission lines.

\subsubsection{\Feii\ subtraction}
\label{subsub:iron}
\begin{figure*}
\begin{center}
\includegraphics[width=0.93\textwidth]{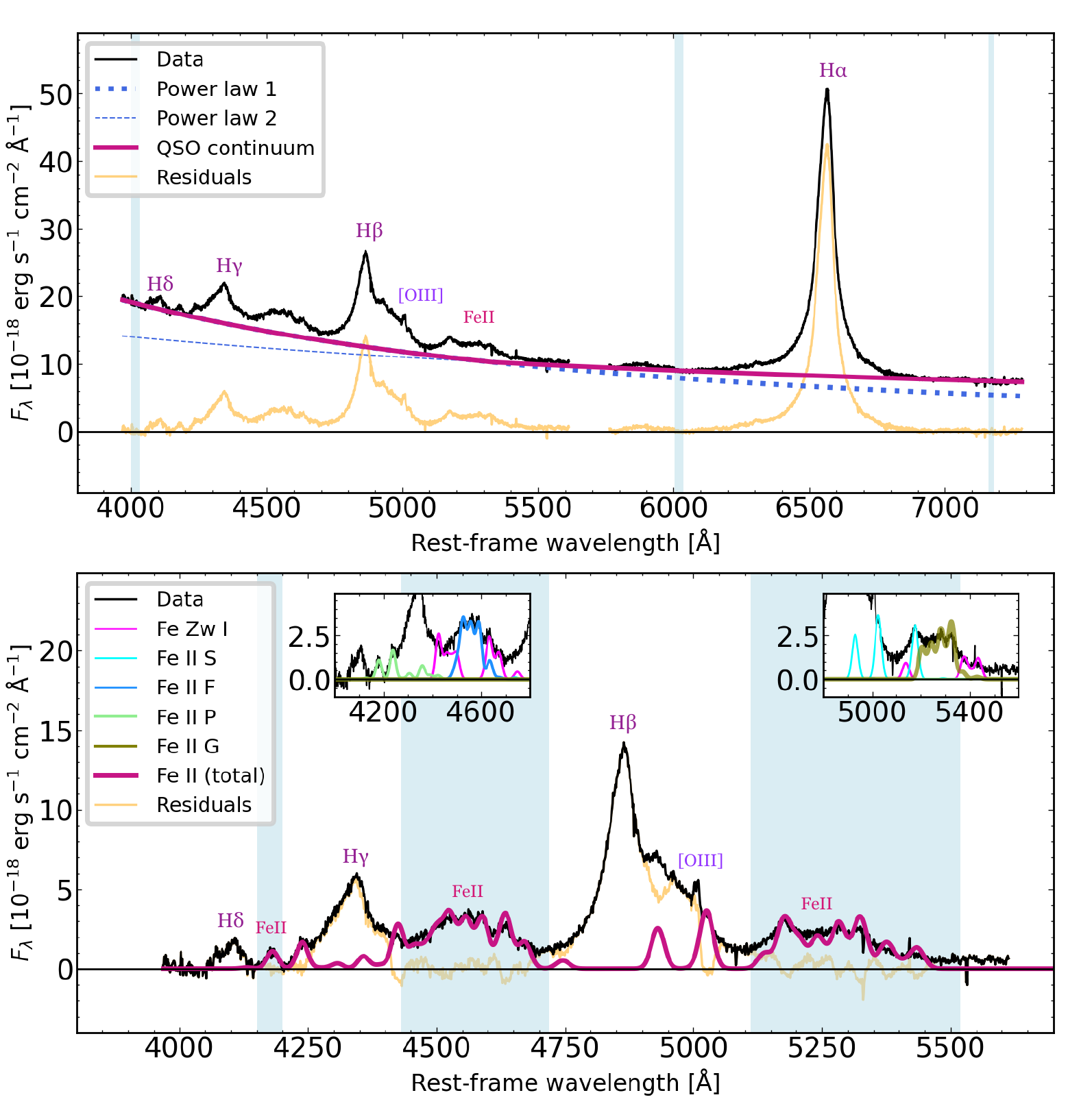}\\
\end{center}
\caption{JWST/NIRSpec rest-frame optical spectrum of the PJ308-21 quasar (black). The spectrum was extracted from a $0.3\arcsec$ aperture and then aperture corrected to account for the PSF broadening with wavelength (see Sect.~\ref{sub:apcor}). Broad Balmer line emission (\Ha , \Hb , \Hg , \Hd ), strong \Feii\ and \Oiii\ emission is clearly seen.
\emph{Top panel:} QSO continuum fit (magenta curve). We highlight the line-free regions used for the continuum fit (pale blue). The orange curve shows the residuals.
\emph{Bottom panel:} \Feii\ line model (magenta). The fitted wavelengths are highlighted in pale blue. We fit the \Feii\ emission using the five multiplets template of \citet{2010ApJS..189...15K} and \citet{2012ApJS..202...10S} (colored lines in the insets). The iron-subtracted spectrum (orange) was then used to fit the Balmer and nebular lines.}
\label{fig_fit_cont}
\end{figure*}

After subtracting the continuum, we model the Fe II multiplets in the spectral range $4000 - 5500\ \AA$. Careful subtraction of the Fe II emission is indeed necessary for deriving an accurate H$\beta$-based black hole mass (see Sect.~\ref{sub:bhmass}), as this line is partially contaminated by iron emission.
We use the tool by \citet{2010ApJS..189...15K} and \citet{2012ApJS..202...10S} to create an \Feii\ template consisting of five groups of lines (labeled as Zw I, S, F, P, G), using reasonable input parameters for an AGN spectrum (i.e. temperature $T = 10000\ K$, velocity width of $2500$ \kms\ and no velocity shift). However, using a different set of input parameters does not change the fit results significantly. In particular, we verify that increasing the \Feii\ width, by smoothing it to match the width of other BLR lines, does not significantly impact the black hole mass derived from the broad \Hb\ full-width at half-maximum (FWHM; see Sect~\ref{sub:bhmass}). We fit the normalization of the five groups (labeled as $I_{\rm ZwI}$, $I_{\rm S}$, $I_{\rm F}$, $I_{\rm P}$, $I_{\rm G}$, see Table~\ref{tab:fitcontiron}) to properly rescale the templates using the spectral regions at $4150-4200$\ \AA , $4430-4720$\ \AA\ and $5110-5520$\ \AA\ (shaded regions in Figure~\ref{fig_fit_cont}, bottom panel). The \Feii\ model is shown in Figure~\ref{fig_fit_cont} (bottom panel), with the iron-subtracted spectrum in orange (see also Table~\ref{tab:fitcontiron} for the fitted parameters). We use the templates of \citet{2010ApJS..189...15K} and \citet{2012ApJS..202...10S}, which secure reliable multiplets structure and intensity. However, the high S/N of our spectrum highlights the limits of the models. Introducing variable widths among the templates due to a velocity stratification of the BLR could possibly improve the fit compared to the simple case here presented.
\begin{figure*}
\begin{center}
\includegraphics[width=1.\textwidth]{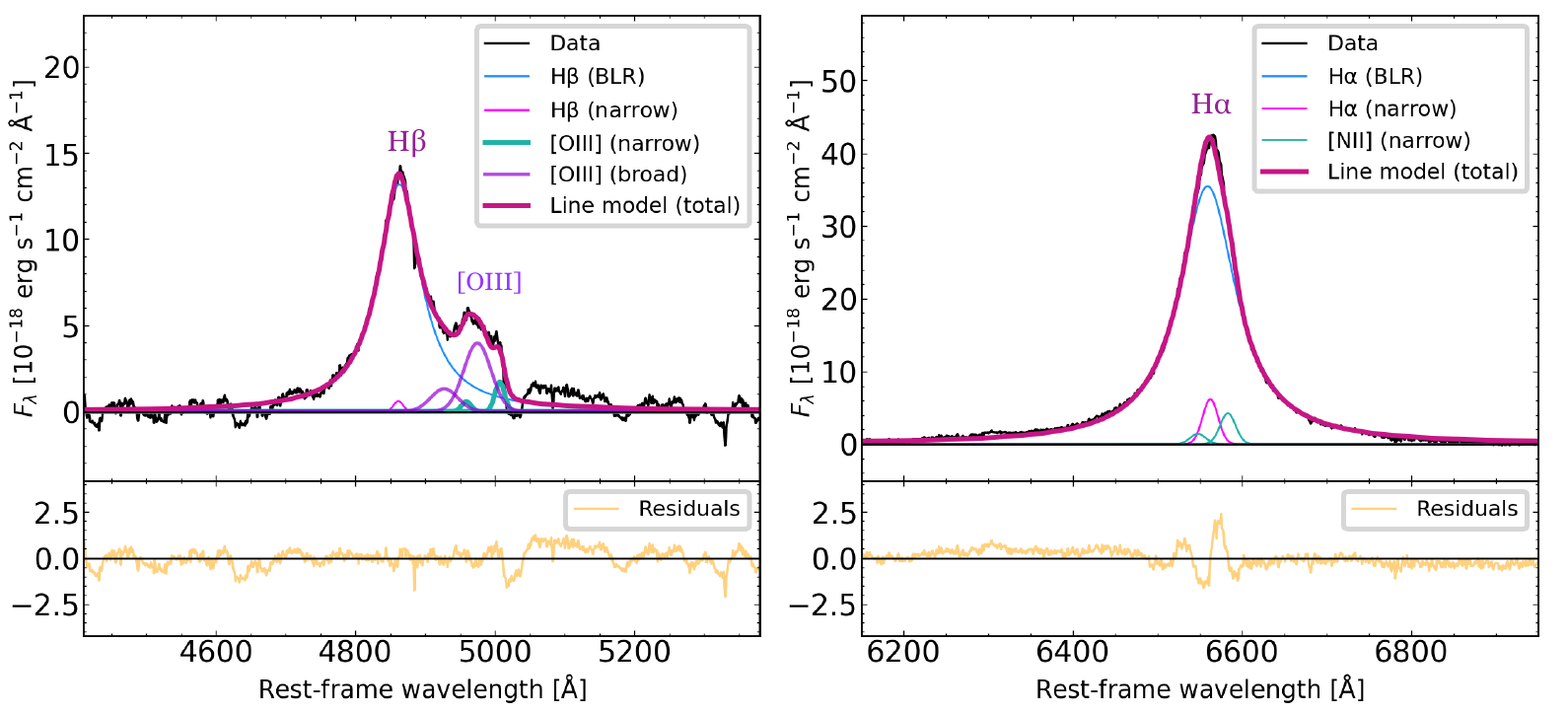}\\
\end{center}
\caption{Fit of the \Hb\ + \Oiii\ (\textit{left}) and \Ha\ + \Nii\ lines (\textit{right}), after the continuum and iron subtraction.
We model the BLR \Hb\ and \Ha\ lines using a Lorentzian function. The nebular lines from the NLR (\Oiii, \Nii, \Ha, \Hb\ narrow components) are each fitted with a Gaussian. We add two further Gaussian components (purple; "broad" components in the text) to fit the emission around 5000 \AA , likely associated with blueshifted \Oiii\ (see also Sect.~\ref{sub:shift} and Figure~\ref{fig_velo}). The residuals are shown in orange.}
\label{fig_fit_lines}
\end{figure*}

\subsubsection{Line fitting}
\begin{table}
\caption{Continuum and iron fitted parameters.}  
\label{tab:fitcontiron}      
\centering                                      
\scalebox{1}{
\begin{tabular}{c c}          
\hline\hline      
Parameter & Value \\[1mm]    
\hline                                 
    $A^{+}$ & $10.28_{-2.21}^{+1.29}$ \\[1mm]
    $\lambda_{\rm break}/\AA$ & $5330.65_{-542.29}^{+1059.70}$ \\[1mm]
    $\alpha_1$ & $2.16_{-0.32}^{+0.65}$ \\[1mm]
    $\alpha_2$ & $1.08_{-0.37}^{+0.04}$ \\[1mm]   
    $I_{\rm ZwI}^{+}$ & $1.28_{-0.03}^{+0.03}$ \\[1mm]    
    $I_{\rm S}^{+}$ & $1.16_{-0.03}^{+0.03}$ \\[1mm]    
    $I_{\rm F}^{+}$ & $0.72_{-0.01}^{+0.01}$ \\[1mm]    
    $I_{\rm P}^{+}$ & $0.59_{-0.05}^{+0.05}$ \\[1mm]    
    $I_{\rm G}^{+}$ & $0.90_{-0.01}^{+0.01}$ \\[1mm]    
\hline 
\end{tabular}}
\tablefoot{The errors correspond to the 16th and 84th percentiles of the MCMC fit. $^{(+)}$The normalization units are $10^{-18}\ \rm erg\ s^{-1} cm^{-2} \AA^{-1}$.}
\end{table}
\begin{table*}
\caption{Emission-line fitted parameters.}  
\label{tab:fitlines}      
\centering                                      
\resizebox*{0.7\textwidth}{!}{ 
\begin{tabular}{c c c c}          
\hline\hline       
Line & Centroid & $\sigma$ & Peak flux \\    
  & ($\AA$, rest-frame)& ($\AA$, rest-frame) & $(10^{-18}\ \rm erg\ s^{-1} cm^{-2} \AA^{-1})$\\
\hline                                   
    H$\alpha$ (BLR) & $6559.53_{-0.15}^{+0.16}$ & $40.91_{-0.28}^{+0.28}$\tablefootmark{*} & $35.46_{-0.31}^{+0.31}$ \\[1mm] 
    H$\beta$ (BLR) & $4862.76_{-0.24}^{+0.25}$ & $37.43_{-0.50}^{+0.51}$\tablefootmark{*} & $13.18_{-0.14}^{+0.14}$ \\[1mm]
    H$\alpha$ (NLR) & $\_$ & tied & $15.41_{-0.81}^{+0.80}$ \\[1mm]
    H$\beta$ (NLR) & $\_$ & tied & $1.52_{-0.51}^{+0.52}$ \\[1mm]
    \Nii\ $\lambda 6548$ (NLR) & $\_$ & tied & tied \\[1mm]
    \Nii\ $\lambda 6584$ (NLR) & $\_$ & tied & $10.60_{-0.66}^{+0.65}$ \\[1mm]
    \Oiii\ $\lambda 4959$ (NLR) & $\_$ & tied & tied \\[1mm]
    \Oiii\ $\lambda 5007$ (NLR) & $\_$ & $6.98_{-0.04}^{+0.02}$ & $4.27_{-0.38}^{+0.38}$ \\[1mm]
    \Oiii\ $\lambda 4959$ (broad) & tied & tied & tied \\[1mm]
    \Oiii\ $\lambda 5007$ (broad) & $4974.74_{-0.64}^{+0.65}$ & $19.60_{-0.52}^{+0.53}$ & $9.93_{-0.23}^{+0.23}$ \\[1mm]
\hline 
\end{tabular}}
\tablefoot{The errors correspond to the 16th and 84th percentiles of the MCMC fit. A "\_" indicates the fixed parameters. \tablefoottext{*}{The reported $\sigma$ is the half-width at half maximum of a Lorentzian function in the functional form $f(\lambda) = f_{\rm peak} \sigma^2/[(\lambda - \lambda_{\rm peak})^2 + \sigma^2]$, where $f_{\rm peak}$ and $\lambda_{\rm peak}$ are the peak flux and centroid respectively.}}
\end{table*}
After subtracting the continuum and \Feii\ model, we fit the line emission. In particular, we focus on the \Ha\ and \Hb\ line complexes, since we use these lines to derive relevant physical quantities (e.g., black hole mass; see Sect.~\ref{sec:results}). We use two Lorentzian functions (one for each line) to model the \Ha\ and \Hb\ lines from the BLR \citep{kollat13}, leaving all parameters (i.e., normalization, width and centroid) free. We also insert six Gaussians to account for the nebular \Oiii\ $\lambda \lambda 4959, 5007$ , \Nii\ $\lambda \lambda 6548, 6584$, \Ha\ and \Hb\ lines from the NLR (referred to as "narrow" components in the rest of the text). As the emission is completely dominated by the broad \Ha\ and \Hb\ lines, we fix the line centroids to the theoretical values to aid the fit, by assuming that the rest-frame velocity of the narrow lines is consistent with the \Cii\ velocity. We also require that all six narrow lines have the same velocity width and fix the ratio of the amplitudes of  \Oiii\ $\lambda 4959$ and \Oiii\ $\lambda 5007$ to the theoretical value (0.33), as well as for the \Nii\ $\lambda 6548$ and \Nii\ $\lambda 6584$ (0.33; e.g., \citealt{osterbrock}). Thus, for the NLR lines, we fit only the normalization of \Oiii\ $\lambda 5007$, \Nii\ $\lambda 6584$, \Ha , \Hb\ and the standard deviation ($\sigma$). Moreover, we add two further Gaussian components to reproduce the emission at $ 4900 - 5100\ \AA$ (the "broad" \Oiii\ components discussed in Sect~\ref{sub:shift}). Under the assumption that this emission is due to \Oiii\ $\lambda \lambda 4959, 5007$ blueshifted with respect to the systemic velocity, we require that the two Gaussians have the same velocity width and offset, and the theoretical ratio for the amplitudes. Therefore, in this case, the free parameters are the centroid, $\sigma$ and the normalization of the \Oiii\ $\lambda 5007$ component.\\
\indent We adopt Maxwellian priors for all the fitted parameters except for the width of the narrow components. In this case we impose that the FWHM can uniformly vary between $50$ and $1000$ \kms\ (a similar approach was adopted by \citealt{vietri18}). The fitted wavelength ranges are $4430 - 5520\ \AA$ and $6160 - 6900\ \AA$.
The fit is shown in Fig.~\ref{fig_fit_lines} (see also Table~\ref{tab:fitlines}).\\
\indent Our multicomponent fit nicely reproduces the line emission for both \Ha\ and \Hb\ regions. As seen for the continuum fitting, the galactic contribution (i.e., NLR emission) is poorly constrained, as the BLR dominates the line emission. There are some residuals at $5000 - 5200\ \AA$, possibly due to uncertainties in the iron modeling described in Sect.~\ref{subsub:iron}. However, this does not significantly impact the \Hb\ fit and the \Hb -based black hole mass estimates. Finally, we note that using a broken power-law rather than a Lorentzian curve to model the \Ha\ and \Hb\ lines from the BLR also gives a consistent black hole mass. The BLR emission may have indeed asymmetric wings that are better reproduced by a broken power-law. A study of these features is however beyond the scope of this work. We also test that adding a broad component for the \Ha\ and \Hb\ lines, as done for the \Oiii\ emission, does not affect the fit, as both the components are in the order of the residuals.

\section{Results}
\label{sec:results}

\subsection{Black hole mass}
\label{sub:bhmass}

\begin{figure*}
\begin{center}
\includegraphics[width=1.\textwidth]{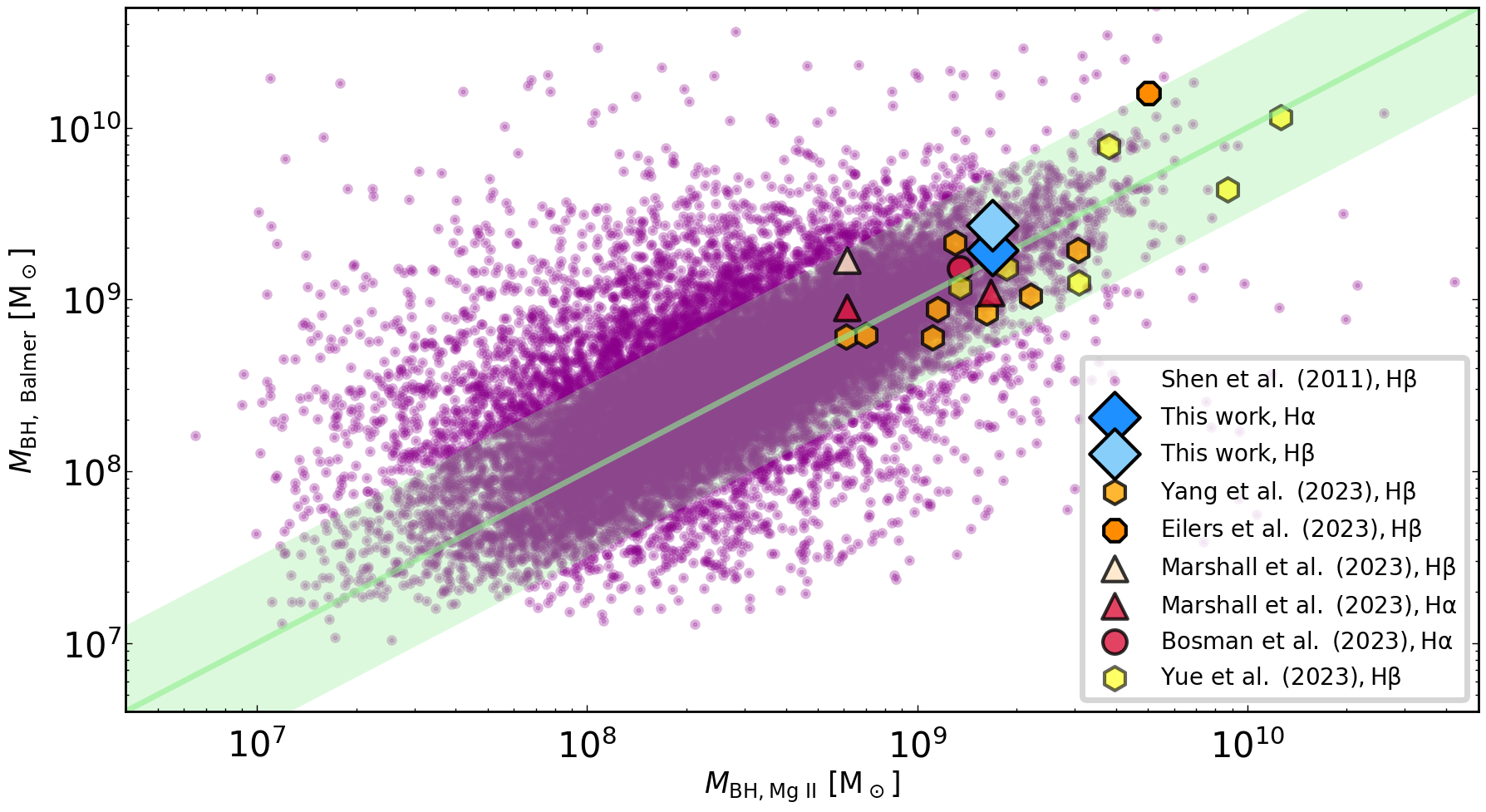}\\
\end{center}
\caption{Black hole masses from SDSS \citep{shen11} and $z > 6$ quasars \citep{bosman23, eilers23, marshall23, yang23, yue23}, with both \Mgii\ and Balmer line based measurements. The green line shows the one-to-one relation. PJ308-21 is indicated with blue diamonds (JWST/NIRSpec and VLT/XSHOOTER data). Overall, the $z > 6$ quasars show nice agreement between the \Mgii\ and \Hb\ or \Ha -based $M_{\rm BH}$, consistent with the $\sim 0.5$ dex scatter of the low-redshift estimates (green region; \citealt{vestpet06}). While a larger sample is needed to confirm these results, these first JWST estimates suggest that the \Mgii\ is a reliable tracer of the black hole mass at both low and high-redshift.}
\label{fig_bhmass}
\end{figure*} 

A common way to estimate the black hole mass in quasars relies on the virial theorem. Assuming that the clouds of the BLR undergo pure gravitational motion around the black hole, we can write the black hole mass as $M_{\rm BH} = R_{\rm BLR} v^2_{\rm BLR} / G$, where $G$ is the gravitational constant, $v_{\rm BLR}$ the clouds' velocity and $R_{\rm BLR}$ their distance from the black hole. Based on reverberation mapping studies in the low-$z$ Universe, the continuum luminosity at 5100 \AA\ is correlated with $R_{\rm BLR}$ (e.g., \citealt{kaspi05}), while the FWHM of the BLR lines (e.g., \Hb , \Ha , \Mgii, \Civ) provide the information for the velocity $v_{\rm BLR}$.
Before the JWST launch, emission lines such as \Mgii\ and, more rarely, \Civ\ were commonly used to measure the black hole masses in high-$z$ quasars (e.g., \citealt{jun15, lira18, onoue19, kaspi21, farina22}). However, the relations that employ these tracers are actually poorly constrained and uncertain, especially for the \Civ\ line \citep{coatman17, marziani19, 2020ApJ...901...55H, zuo20}. It is thus not clear if the black hole masses based on these lines are reliable. On the other hand, \Hb\ has been extensively studied to derive empirical calibrations for black hole mass at low redshift (e.g., \citealt{bentz09, derosa18}; see \citealt{shen13} for a review), and it is thus considered the most robust tracer of this physical quantity. After the launch of JWST, we can now use this line for measuring the black hole mass even at high-redshift (see also \citealt{ding23, eilers23, marshall23, yang23}).\\
\indent We estimate the black hole mass of PJ308-21 using the relation of \citet{vestpet06}, which uses the $5100\ \AA$ QSO continuum luminosity $L_{\lambda}(5100\ \AA)$ and the FWHM of the \Hb : 

\small
\begin{equation}
 \log \left( \dfrac{M_{\rm BH, H\beta}}{M_{\odot}} \right) = 2 \log \left( \dfrac{\rm FWHM(H\beta)}{1000\ \rm km\ s^{-1}} \right) \ + \ 0.5 \log \left( \dfrac{\lambda L_{\lambda}(5100 \AA)}{10^{44} \rm erg\ s^{-1}} \right) + 6.91
 \label{eq:hb}
\end{equation}
\normalsize

\noindent We estimate the continuum monochromatic luminosity $L_{\lambda}(5100\ \AA) = (4.9^{+0.4}_{-0.2}) \times 10^{42}\ \rm erg\ s^{-1} \AA^{-1}$ from the power law value at 5100 \AA\ while the $\rm FWHM(H\beta)$ is $\rm  4616^{+63}_{-61}\ km\ s^{-1}$. The black hole mass is:
\begin{equation*}
M_{\rm BH, H\beta} = (2.7 \pm 0.2) \times 10^{9}\ \rm M_{\odot}
\end{equation*}
We note that the uncertainties on the black hole mass due to random errors from the MCMC fit are around the $7 \%$. However, the adopted calibration has an intrinsic scatter of $\sim 0.5\ \rm dex$, which thus represents the main source of uncertainty \citep{vestpet06}.\\
\indent \citet{2005ApJ...630..122G} found that there is a correlation between the 5100 \AA\ continuum luminosity and the \Ha\ luminosity (see their Eq. 1). A correlation is also present between the \Hb\ and \Ha\ FWHM (see their Eq. 3; see also \citealt{shen11}, Eq. 9). Using these correlations and Eq.~\ref{eq:hb} it is thus possible to obtain a calibration for the black hole mass with the \Ha\ line \citep{shen11}:
\small
\begin{equation}
 \log \left( \dfrac{M_{\rm BH, H\alpha}}{M_{\odot}} \right) = 0.379 \ + \ 0.43 \log \left( \dfrac{L(\rm H\alpha)}{10^{42}\ \rm erg\ s^{-1}} \right) \ + \ 2.1 \log \left( \dfrac{\rm FWHM(H\alpha)}{\rm km\ s^{-1}} \right) 
\label{eq:ha} 
\end{equation}
\normalsize
where $L(\rm H\alpha)$ and $\rm FWHM (\rm H\alpha)$ are the luminosity and FWHM of the \Ha\ broad component. We estimate the former quantity $ L(\rm H\alpha) = (1.84 \pm 0.01) \times 10^{45}\ erg\ s^{-1}$ by integrating the \Ha\ Lorentzian component from $6300\ \AA$ to $6800\ \AA$. 
The FWHM of the Lorentzian is $\rm FWHM (\rm H\alpha) = 3739\pm 26\ km\ s^{-1}$. 
The resulting \Ha -based black hole mass is
\begin{equation*}
M_{\rm BH, H\alpha} = (1.93 \pm 0.03) \times 10^{9}\ \rm M_{\odot}
\end{equation*}
We find that the $M_{\rm BH, H\beta}$ and $M_{\rm BH, H\alpha}$ values agree with each other within a factor of $\lesssim 1.5$ (see also Table~\ref{table:phys}), as expected from the intrinsic scatter of the adopted correlations.\\
\indent We can compare our estimates of the black hole mass with the Mg II-based measurement for PJ308-21. Based on the relation of \citet{shen11}, \citet{farina22} found
\begin{equation*}
M_{\rm BH, MgII} = (2.65^{+0.32}_{-0.56}) \times 10^{9}\ \rm M_{\odot}
\end{equation*}
This value is consistent with the \Hb\ estimate.
For this quasar \citet{farina22} also derived a \Civ\ based black hole mass using the relation of \citet{coatman17}, i.e., $M_{\rm BH,C IV} = (2.09^{+0.52}_{-0.49}) \times 10^{9}\ \rm M_{\odot}$, which is consistent with the estimates from the other lines.\\ 
\indent Figure~\ref{fig_bhmass} compares the PJ308-21 black hole mass estimates with estimates from SDSS \citep{shen11} and the other $z > 6$ quasars with both \Mgii\ and \Hb\ or \Ha\ black hole mass measurements from JWST data (\citealt{bosman23, eilers23, marshall23, yang23, yue23}). 
The \Mgii\ masses are based on the \citet{vo09} and \citet{shen11} calibration (see \citealt{pons19, reed19, shen19, wang21, yang21, farina22, mazzucchelli23}) while the \Hb\ and \Ha\ measurements are based on \citet{vestpet06} and \citet{2005ApJ...630..122G}. 
Overall, we see that the $z > 6$ QSOs lie well inside the $\sim 0.5$\ dex scatter of the low-redshift estimates. Despite larger sample are necessary to constrain $z > 6$ black hole masses via Balmer lines targeting also less massive black holes (i.e., $M_{\rm BH} \sim 10^8 \rm M_{\odot}$), these first findings suggest that the \Mgii\ is a reliable mass tracer in distant quasars.


\begin{table}
\caption{Physical quantities of PJ308-21 derived from line fitting.}      
\label{table:phys}      
\centering                                      
\scalebox{0.9}{
\begin{tabular}{c c}          
\hline\hline       
PJ308-21 &  \\    
\hline                                   
$L_{\lambda}(5100\ \AA)$ & $(4.9^{+0.4}_{-0.2}) \times 10^{42}\ \rm erg\ s^{-1} \AA^{-1}$\\
$L(\rm H\alpha)$ & $(1.84 \pm 0.01) \times 10^{45}\ \rm erg\ s^{-1}$\\
    $\rm FWHM(H\beta)$ & $\rm 4616^{+63}_{-61}\ km\ s^{-1}$\\
    $\rm FWHM(H\alpha)$ & $\rm 3739 \pm 26\ km\ s^{-1}$\\ 
    $\rm FWHM([O\ III])$ & $ 2776^{+75}_{-74}$ \kms\\
    $\Delta v_{\rm H\beta}$ & $\rm 87 \pm 15$ \kms \\    
    $\Delta v_{\rm H\alpha}$ & $\rm -149 \pm 7$ \kms \\    
    $\Delta v_{\rm [O\ III]}$ & $-1922 \pm 39$ \kms\\   
    $M_{\rm BH, H\beta}$ & $ (2.7 \pm 0.2) \times 10^{9}\ \rm M_{\odot}$\\
    $M_{\rm BH, H\alpha}$ & $ (1.93 \pm 0.03) \times 10^{9}\ \rm M_{\odot}$\\
    $L_{\rm bol}$ & $(2.3^{+0.2}_{-0.1}) \times 10^{47} \rm erg\ s^{-1}$\\
    $\lambda_{\rm Edd, H\beta}$ &  $0.67^{+0.12}_{-0.05}$\\
    $\lambda_{\rm Edd, H\alpha}$ & $0.96^{+0.10}_{-0.06}$\\
    $EW_{\rm [O\ III]}$ & $19 \pm 2\ \AA$\\ 
    $EW_{\rm H\beta}$ & $118^{+8}_{-11}\ \AA$\\    
    $EW_{\rm Fe\ II}$ & $43^{+2}_{-3}\ \AA$\\  
    $R_{\rm Fe\ II}$ & $0.36^{+0.04}_{-0.06}$\\            
\hline                                             
\end{tabular}
}
\tablefoot{We report the random errors from the MCMC fit. The statistical uncertainty, due to object-to-object differences, on the black hole mass estimates is $\sim 0.5$ dex \citep{vestpet06} and affects also the Eddington ratios.}
\end{table}

\subsection{Eddington ratio}
\label{sub:edd}
We use the black hole mass to evaluate the Eddington ratio $\lambda_{\rm Edd} =  L_{\rm bol}/L_{\rm Edd} $, i.e., the ratio between the quasar bolometric luminosity $L_{\rm bol}$ and the Eddington luminosity $L_{\rm Edd}$ (see \citealt{peterson97}). Assuming spherical geometry, the latter quantity can be written as:
\begin{equation*}
L_{\rm Edd} = \dfrac{4 \pi G M_{\rm BH} m_{\rm p} c}{\sigma_{\rm T}} = 1.257 \times 10^{38}\ \left( \dfrac{M_{\rm BH}}{M_{\odot}} \right) \rm erg\ s^{-1}
\end{equation*}
where $\sigma_{\rm T}$ is the cross-section for Thomson scattering, $m_{\rm p}$ is the proton mass and $c$ is the light speed in vacuum. 
Based on the $M_{\rm BH, H\beta}$ estimate, 
we find for PJ308-21 $L_{\rm Edd, H\beta} = (3.4^{+0.3}_{-0.1}) \times\ 10^{47}\ \rm erg\ s^{-1}$ ($L_{\rm Edd, H\alpha} = (2.43 \pm 0.04) \times\ 10^{47}\ \rm erg\ s^{-1}$).
The bolometric luminosity $L_{\rm bol}$ can be estimated from the monochromatic luminosity $L_{\lambda}(5100 \AA)$ by assuming a bolometric correction. We use the relation from \citet{richards06}, updated by \citet{shen11}:
\begin{equation*}
L_{\rm bol} = 9.26 \times \lambda L_{\lambda}(5100 \AA)
\end{equation*}
based on 259 quasars from SDSS. This gives a QSO bolometric luminosity of $L_{\rm bol} = (2.3^{+0.2}_{-0.1}) \times 10^{47}\ \rm erg\ s^{-1}$. Therefore, the Eddington ratio for  PJ308-21 based on the \Hb\ line is $\lambda_{\rm Edd, H\beta} = 0.67^{+0.12}_{-0.05}$ ($\lambda_{\rm Edd, H\alpha} = 0.96^{+0.10}_{-0.06}$ for \Ha ).
We note that adopting a different bolometric correction would imply lower Eddington ratio, in the range $0.32 < \lambda_{\rm Edd, H\beta} < 0.47$ ($0.45 < \lambda_{\rm Edd, H\alpha} < 0.66$ for \Ha ; \citealt{shen11, krawczyk13, saccheo23}). We note also that the true uncertainty on $\lambda_{\rm Edd}$ is likely higher due to the statistical uncertainty, due to object-to-object differences, on the black hole mass calibrations and on the bolometric luminosity.\\
\indent We can look at the high Eddington ratio of PJ308-21 in the context of the $z \sim 6$ QSO population (see also \citealt{yang23}). Based on VLT/XSHOOTER spectra of 38 QSO at $5.8 < z < 7.5$, \citet{farina22} found that these objects show higher values of $\lambda_{\rm Edd}$ than the low-redshift population at fixed $L_{\rm bol}$ \citep{shen11}. This suggests that in the early Universe QSOs tended to accrete at higher rates than the low-$z$ AGN, even if exceptions are present among low-luminosity quasars \citep{kim18, matsuoka19}. Besides, this trend can be affected by observational biases toward a selection of more massive black holes and higher bolometric luminosities (and thus higher Eddington ratios) at high redshift.
PJ308-21 has an Eddington ratio based on \Hb\ and \Ha\ above, but still consistent (given the statistical uncertainties in the black hole masses) with the median value $\lambda_{\rm Edd}^{\rm med} = 0.48^{+0.06}_{-0.02}$ of the $5.8 < z < 7.5$ quasar sample of \citet{farina22}. The \Hb\ value is also in perfect agreement with the \Mgii -based $\lambda_{\rm Edd} = 0.66^{+0.17}_{-0.07}$ for PJ308-21 measured by \citet{farina22} adopting the bolometric correction for $3000\ \AA$\ of \citet{richards06} updated by \citet{shen11}.  

\subsection{Correlations between \Feii , \Oiii\ and \Hb}
\label{sub:corr}
\begin{figure}
\begin{center}
\includegraphics[width=0.52\textwidth]{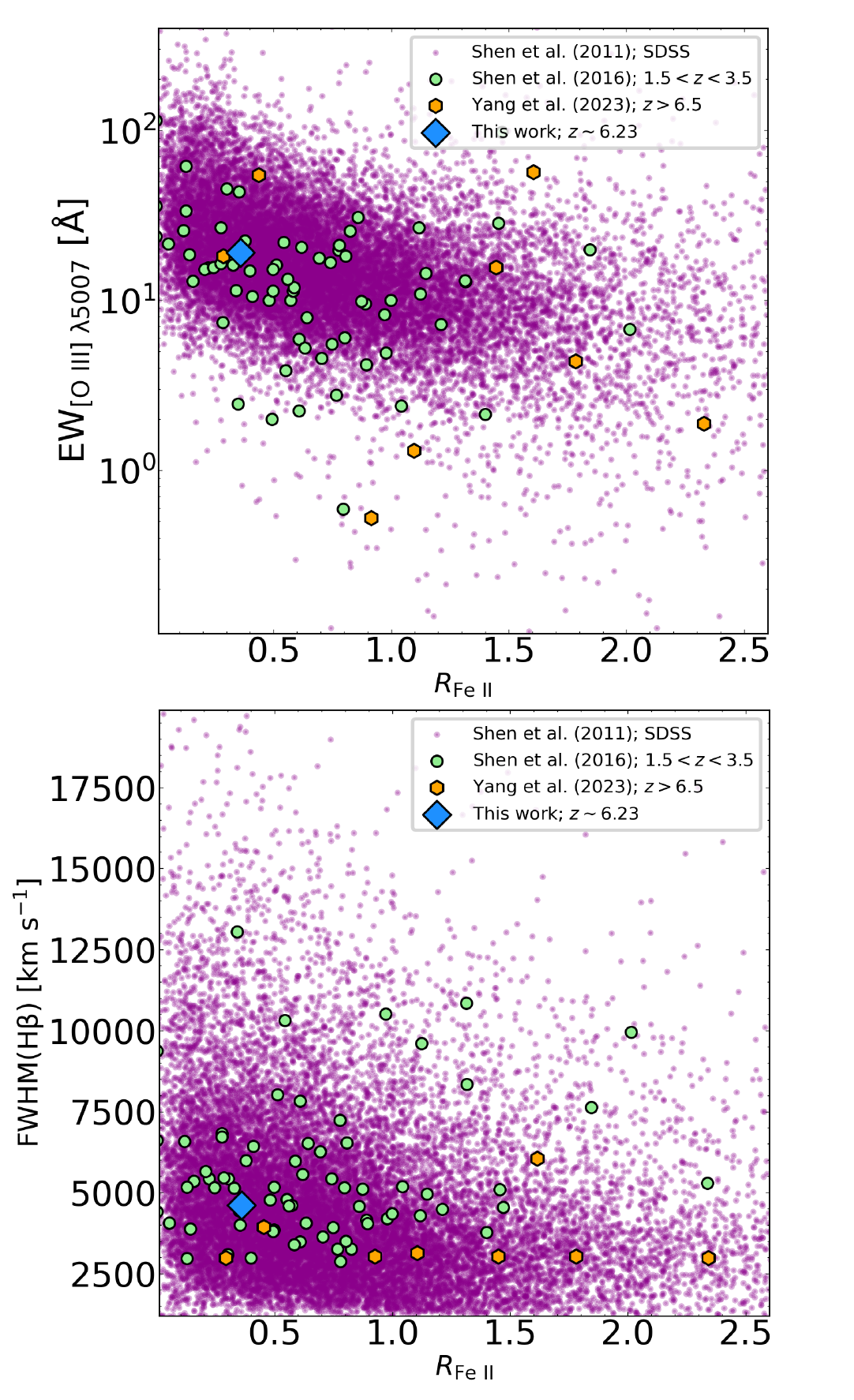}\\
\end{center}
\caption{Anti-correlations between the \Feii\ strength, parametrized by $R_{\rm Fe\ II}$, the \Oiii\ $EW$ and \Hb\ FWHM. \textit{Top  panel}: the anti-correlation between $R_{\rm Fe\ II}$ and $EW_{\rm [O\ III]}$ (EV1 relation). PJ308-21 (blue diamond) shows a value in agreement with both low-redshift \citep{shen11}, $1.5 < z < 3.5$ \citep{shen16} and $z > 6$ quasars \citep{yang23}. \textit{Bottom panel}: anti-correlation between $R_{\rm Fe\ II}$ and FWHM(\Hb ) (2D EV1). Also in this case, the PJ308-21 value appears consistent with the findings of other studies.}
\label{fig_rfeii}
\end{figure}
As initially shown by \citet{boroson92}, the strength of the \Feii\ and \Oiii\ $\lambda 5007$ emission is anti-correlated in low-redshift quasars. The so-called "Eigenvector 1 correlation" ($EV1$) was then confirmed by other papers (e.g., \citealt{sulentic04, sulentic06, 2010ApJS..189...15K, runnoe13, pennell17}), comparing the \Feii\ strength, quantified by the ratio between the \Feii\ and \Hb\ rest-frame equivalent width, i.e., $R_{\rm Fe\ II} = EW_{\rm Fe\ II}/EW_{\rm H\beta }$, and the \Oiii\ rest-frame equivalent width, $EW_{\rm [O\ III]}$. The physical origin of this relation is still poorly understood, although it may be related to the processes governing the black hole accretion \citep{shen16}.  
At higher redshifts, \citet{shen16} found that the anti-correlation is in place in $1.5 < z < 3.5$ QSOs, with slightly lower $EW_{\rm [O\ III]}$ compared to low redshift (see also \citealt{sulentic04, sulentic06}). Before the JWST launch, it was not possible to investigate this relation for quasars at $z > 6$. First results using slitless JWST/NIRCam spectroscopy in eight $z > 6$ QSOs by \citet{yang23} are consistent with low-redshift samples.\\
\indent For PJ308-21 we find an \Oiii\ $\lambda 5007$ rest-frame equivalent width $EW_{\rm [O\ III]} = 19 \pm 2$ \AA . We measure this quantity including both the \Oiii\ $\lambda 5007$ narrow and broad component, following \citet{shen11}. The $EW$ of the \Hb\ broad component is $EW_{\rm H\beta } = 118^{-8}_{-11}$ \AA\ while the $EW_{\rm Fe\ II}$ is $43^{+2}_{-3}$ \AA . The latter quantity was obtained by directly integrating the continuum-subtracted \Feii\ emission in the range $4434-4684$ \AA\ \citep{shen11}.
We integrate the data rather than the \Feii\ line fit in order to avoid uncertainties in the iron modeling (see Sect.~\ref{subsub:iron} and Fig.~\ref{fig_fit_cont}, bottom panel). The resulting $R_{\rm Fe\ II}$ is $R_{\rm Fe\ II} = 0.36^{+0.04}_{-0.06}$. We note that the iron emission may be contaminated by the blue/red wing of the \Hb / \Hg\ BLR lines. We neglect this contribution when estimating the iron $EW$ for consistency with \citet{shen11}. We also test that subtracting the \Hb\ and \Hg\ emission does not significantly affect any of the results about the EV correlations for the quasar. In that case we obtain indeed $EW_{\rm Fe\ II}$ is $36^{+2}_{-3}$ \AA\ and $R_{\rm Fe\ II} = 0.31^{+0.03}_{-0.05}$, which place PJ308-21 in a similar location in the two panels of Figure~\ref{fig_rfeii}.\\ 
\indent In Figure~\ref{fig_rfeii} (top panel), we compare PJ308-21 with the DR7 SDSS quasars from \citet{shen11} and a sample of 74 luminous ($L_{\rm bol} = 10^{46.2-48.2}\ \rm erg\ s^{-1}$) $1.5 < z < 3.5$ QSOs \citep{shen16}. We show also the eight $z > 6$ quasars from \citet{yang23}.
Our source is consistent with the values presented by both low-redshift and high-redshift quasars, showing a low $R_{\rm Fe\ II}$ corresponding to a high $EW_{\rm [O\ III]}$.\\
\indent We also explore the anticorrelation between $R_{\rm Fe\ II}$ and the \Hb\ FWHM ("2D EV1", e.g., \citealt{boroson92, shenHo14}); see Figure~\ref{fig_rfeii} (bottom panel). 
Compared to low redshift objects, high-redshift quasars show a systematic offset in the \Hb\ FWHM up to $z >6$ \citep{shen16, yang23}. This is due to a selection effect as at high redshift we mainly sample the most massive black holes, which correspond to larger \Hb\ FWHM. 
We find that, for this relation as well, PJ308-21 shows values in line with the findings for the DR7 SDSS quasars and higher-redshift QSOs. 


\subsection{Velocity shifts}
\label{sub:shift}
Several papers have studied the bulk velocity shifts of the BLR gas (e.g., \citealt{gaskell82, richards02, meyer19, onoue20}). In particular, high-ionization lines like \Civ\ $\lambda 1549$ and \Siiv\ $\lambda 1397$ have significantly higher velocity shifts relative to low-ionization broad lines, such as \Mgii\ and Balmer lines. These velocity offsets are thought to be due to outflowing clouds from the BLR driven by winds or X-ray radiation \citep{murray95}.\\ 
\indent According to \citet{schindler20}, PJ308-21 displays a \Civ\ blueshifted emission of $-2823^{+200}_{-188}$ \kms\ with respect to the \Cii\ systemic velocity. On the other hand, the low ionization \Mgii\ line shows a blueshift of $-167^{+98}_{-101}$ \kms . From the NIRSpec spectrum, we find that the \Ha\ emission is blueshifted of $\Delta v_{\rm H\alpha} = -149 \pm 7$ \kms while the \Hb\ line is redshifted of $\Delta v_{\rm H\beta} = 87 \pm 15$ \kms . \citet{shen11} found no significant offset in \Ha\ and \Hb\ relative to the narrow \Oiii\ and \Sii\ lines in their SDSS DR7 sample of quasars. The overall small shifts found for the \Mgii\ and Balmer lines in PJ308-21 suggest that the clouds these lines originate from are mostly dominated by virial motion around the black hole rather than outflowing material \citep{gaskell13}, which is a required assumption for the validity of the single-epoch black hole mass measurements presented in Sect~\ref{sub:bhmass}.
\begin{figure}
\begin{center}
\includegraphics[width=0.47\textwidth]{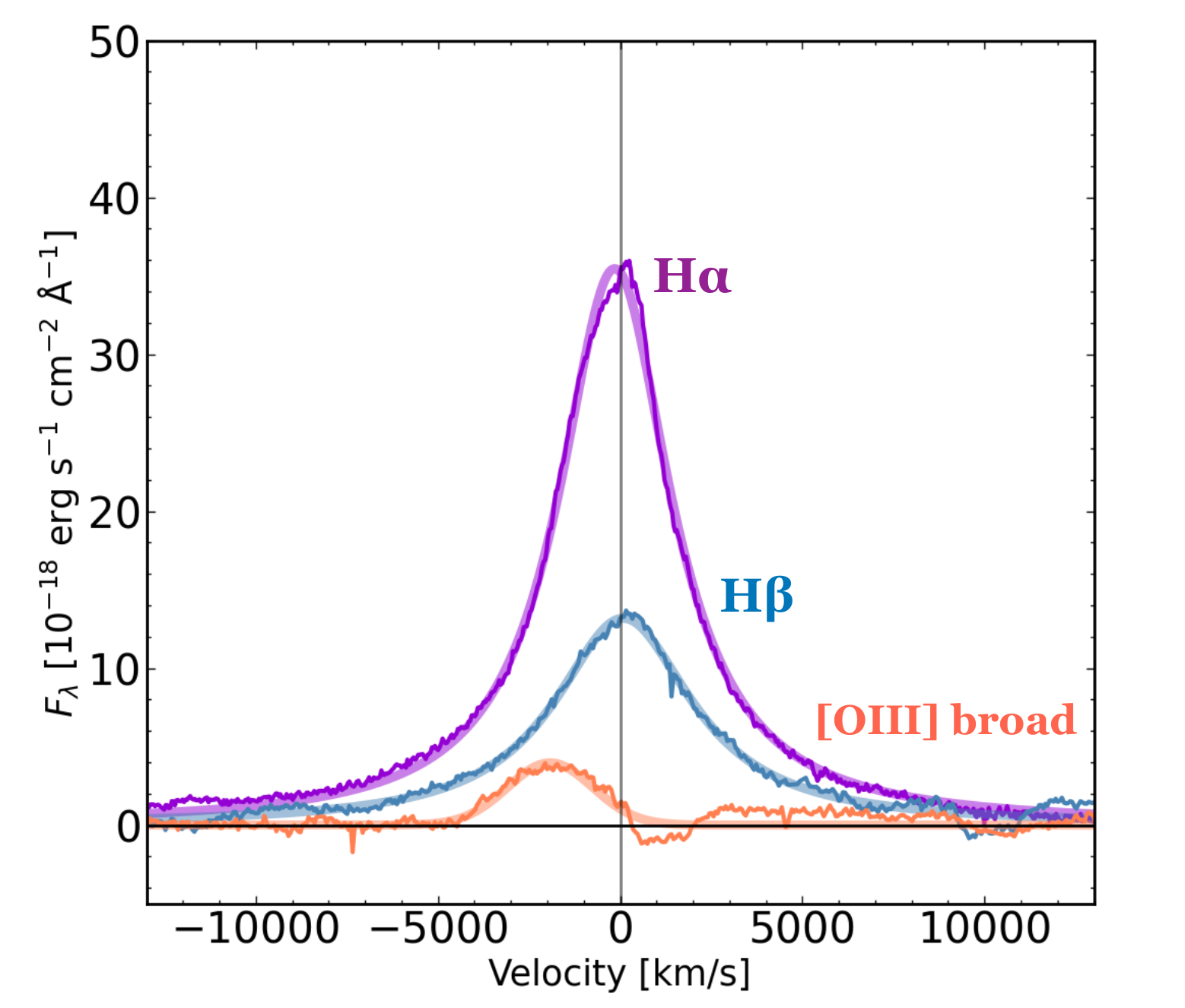}\\
\end{center}
\caption{Emission from the \Ha\ and \Hb\ BLR lines and the \Oiii\ $\lambda 5007$ broad component as a function of the velocity. Both the subtracted-data (solid lines) and the fitted models (semi-transparent thick curves) are shown. The systemic velocity, set by the vertical grey line, is based on the \Cii\ redshift. The \Ha\ and \Hb\ lines are offset of a small amount compared to the zero-reference point ($\Delta v_{\rm H\alpha} = -149 \pm 7$ \kms\ and $\Delta v_{\rm H\beta} = 87 \pm 15$ \kms ).
On the other hand, we see that the \Oiii\ $\lambda 5007$ broad component shows a velocity blueshift of $\Delta v_{\rm [O\ III]} = -1922 \pm 39$ \kms\ and a FWHM(\Oiii) = $2776^{+75}_{-74}$ \kms . This blueshifted emission highlights possible outflowing material from the QSO host galaxy.}
\label{fig_velo}
\end{figure}
We also estimate the velocity shift of the \Oiii\ broad components, which is commonly used as a tracer of ionized outflows (e.g., \citealt{cano12, harrison12, harrison14, carniani15, zakamska16, loiacono19, tozzi21}). These outflows are often found to extend to kpc-scales, well beyond the nuclear region, and thus can affect the host galaxy and its gas content. As discussed in Sect.~\ref{sub:fit}, we fitted the \Oiii\ emission using four Gaussian functions, two narrow and two broad, with the former centered on the systemic velocity of the system (Figure~\ref{fig_fit_lines}, left). The \Oiii\ $\lambda 5007$ broad line is blueshifted by $\Delta v_{\rm [O\ III]} = -1922 \pm 39$ \kms\ relative to the systemic velocity traced by \Cii . The FWHM of the same line is FWHM(\Oiii) = $2776^{+75}_{-74}$ \kms . While this suggests significant outflowing ionized gas from the NLR of the quasar (see Figure~\ref{fig_velo}), we highlight that the \Oiii\ modeling in this source is quite challenging, as this emission is strongly blended with the \Hb\ and iron lines from the BLR. This makes the errors on $\Delta v_{\rm [O\ III]}$ and FWHM(\Oiii) likely underestimated.
Strong blueshifted \Oiii\ emission was also found in two out of $z > 6.5$ quasars studied by \citet{yang23}. In their case, the velocity offsets are lower at $\sim -630$ \kms\ and $\sim -1690$ \kms , and FWHM$\sim 1461$ \kms\ and $\sim 3805$ \kms .\\ 
\indent High-velocity outflows in high-$z$ sources are predicted by models to play a role in regulating both the black hole and galaxy growth (e.g., \citealt{dimatteo05, hopkins16, weinberger17, costa22, lupi22}). 
At intermediate and low redshift, several studies using optical/NIR, far-infrared and mm/sub-mm facilities assessed the role of outflows in shaping the AGN host galaxies (e.g., \citealt{feruglio10, sturm11, liu13, zakamska14, brusa15, cresci15, nardini15, bischetti17, fiore17, gonzalez17}).
Outflows have been detected in only a small number of $z\sim 6$ quasars \citep{stanley19, bischetti22, butler23, salak24} and in some cases their presence is ambiguous \citep{maiolino12, cicone15, meyer22}.
The first JWST results suggest the existence of strong \Oiii\ outflows at $z > 6$ for the first time, demonstrating the capabilities of JWST in detecting these features at the end of Reionization. With the uncertainties in mind, PJ308-21 shows remarkably large velocity offset and FWHM of the \Oiii\ emission, significantly higher than typical values at lower redshift \citep{liu13, rodz13, brusa15b, perna15, harrison16, bischetti17}.\\ 
\indent Unfortunately, we cannot estimate the size of the outflowing \Oiii\ component from the IFU data as it is outshone by the QSO emission. This quantity would enable indeed an estimate of the mass outflow rate, i.e., the amount of ionized gas expelled per time unit (e.g., \citealt{cano12, carniani15}). However, after PSF subtraction, Decarli et al. (in prep.) find a possible redshifted component located above the host galaxy, extending to the Northwest with much lower velocity ($\sim 300$ \kms ) which may trace the receding cone of the outflow and can be related to the Lyman $\alpha$ nebula enshrouding the quasar \citet{farina22}. Farina et al. (in prep.) will discuss this component in greater detail.    
\section{Conclusions}
\label{sec:concl}

We presented the JWST/NIRSpec IFU spectrum of the $z \sim 6.23$ quasar PJ308-21. The NIRSpec dataset shows the rest-frame optical emission in the $3700 - 7300$ \AA\ range. We briefly summarize the main results of this paper:\\

i) We estimate the black hole mass using the \Hb\ and \Ha\ broad lines. We find that PJ308-21 harbors a mature black hole, with $M_{\rm BH, H\beta} = (2.7 \pm 0.2) \times 10^{9}\ \rm M_{\odot}$. The \Ha -based black hole mass is $M_{\rm BH, H\alpha} = (1.93 \pm 0.03) \times 10^{9}\ \rm M_{\odot}$. The two values are consistent within a factor $\lesssim 1.5$, within the $\sim 0.5$ dex statistical uncertainties of the mass calibrations. The \Hb\ value is also consistent with a previous estimate of the black hole mass $M_{\rm BH, Mg\ II} = (2.65^{+0.32}_{-0.56}) \times 10^{9}\ \rm M_{\odot}$ derived by \citet{farina22} using the \Mgii\ line. All the estimates are consistent with the \Civ\ based black hole mass by \citet{farina22}.\\

ii) PJ308-21 shows a high Eddington ratio $\lambda_{\rm Edd, H\beta} = 0.67^{+0.12}_{-0.05}$ ($\lambda_{\rm Edd, H\alpha} = 0.96^{+0.10}_{-0.06}$), in line with the population of $z \gtrsim 6$ QSOs (see \citealt{farina22, yang23}).\\

iii) PJ308-21 shows values of $\rm EW_{\rm [O\ III]}$ and $R_{\rm Fe\ II}$ similar to those of low-$z$ \citep{shen11} and intermediate redshift quasars ($1.5 < z < 3.5$, \citealt{shen16}). The same holds for the \Hb\ FWHM and $R_{\rm Fe\ II}$ anticorrelation ("2D EV1").\\

iv) We find evidence of a blueshifted and broad \Oiii\ component, with a velocity shift $\Delta v_{\rm [O\ III]} = -1922 \pm 39$ \kms\ relative to the \Cii -based systemic velocity, and a FWHM(\Oiii) = $2776^{+75}_{-74}$ \kms . Both the FWHM and $\Delta v_{\rm [O\ III]}$ are larger than in quasars at lower redshift. Despite the uncertainties affecting the line fitting, this could be one of the first detections of an \Oiii\ outflow from a $z \gtrsim 6$ quasar.\\

Our study demonstrates the unique capabilities of JWST/NIRSpec in dissecting the emission from quasars at $z \gtrsim 6$ with modest amounts of telescope time ($\sim 1.5\ \rm hrs$ on source in our case). Future NIRSpec campaigns on larger samples will be crucial to unveiling new and unparalleled insights on the quasar population at cosmic dawn.

\begin{acknowledgements} 
We thank the referee for their useful comments. LF and RD acknowledge support from the INAF GO 2022 grant "The birth of the giants: JWST sheds light on the build-up of quasars at cosmic dawn". LB acknowledges support from NSF award 2307171.
SEIB is supported by the Deutsche Forschungsgemeinschaft (DFG) under Emmy Noether grant number BO 5771/1-1. RAM acknowledges support from the Swiss National Science Foundation (SNSF) through project grant 200020\_207349. AP acknowledges support from Fondazione Cariplo grant no. 2020-0902. JTS is supported by the Deutsche Forschungsgemeinschaft (DFG, German Research Foundation) - Project number 518006966. BT acknowledges support from the European Research Council (ERC) under the European Union's Horizon 2020 research and innovation program (grant agreement number 950533), and the Israel Science Foundation (grant number 1849/19). MT acknowledges support from the NWO grant 016.VIDI.189.162 (“ODIN”). MV gratefully acknowledges financial support from the Independent Research Fund Denmark via grant number DFF 8021-00130.
This work is based on observations made with the NASA/ESA/CSA James Webb Space Telescope. The data were obtained from the Mikulski Archive for Space Telescopes at the Space Telescope Science Institute, which is operated by the Asssociation of Universities for Research in Astronomy, Inc., under NASA contract NAS 5-03127 for JWST. These observations are associated with the GO program 1554.
\end{acknowledgements}

\bibliographystyle{aa}
\bibliography{main} 




\begin{appendix}
\section{Aperture correction}
\label{app:aper}
\begin{figure*}
\begin{center}
\includegraphics[width=0.45\textwidth]{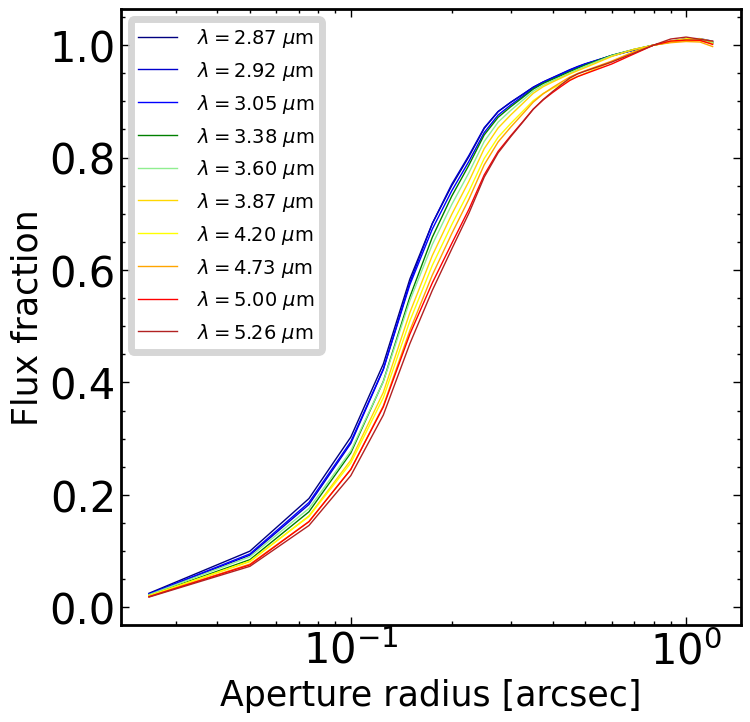}
\includegraphics[width=0.467\textwidth]{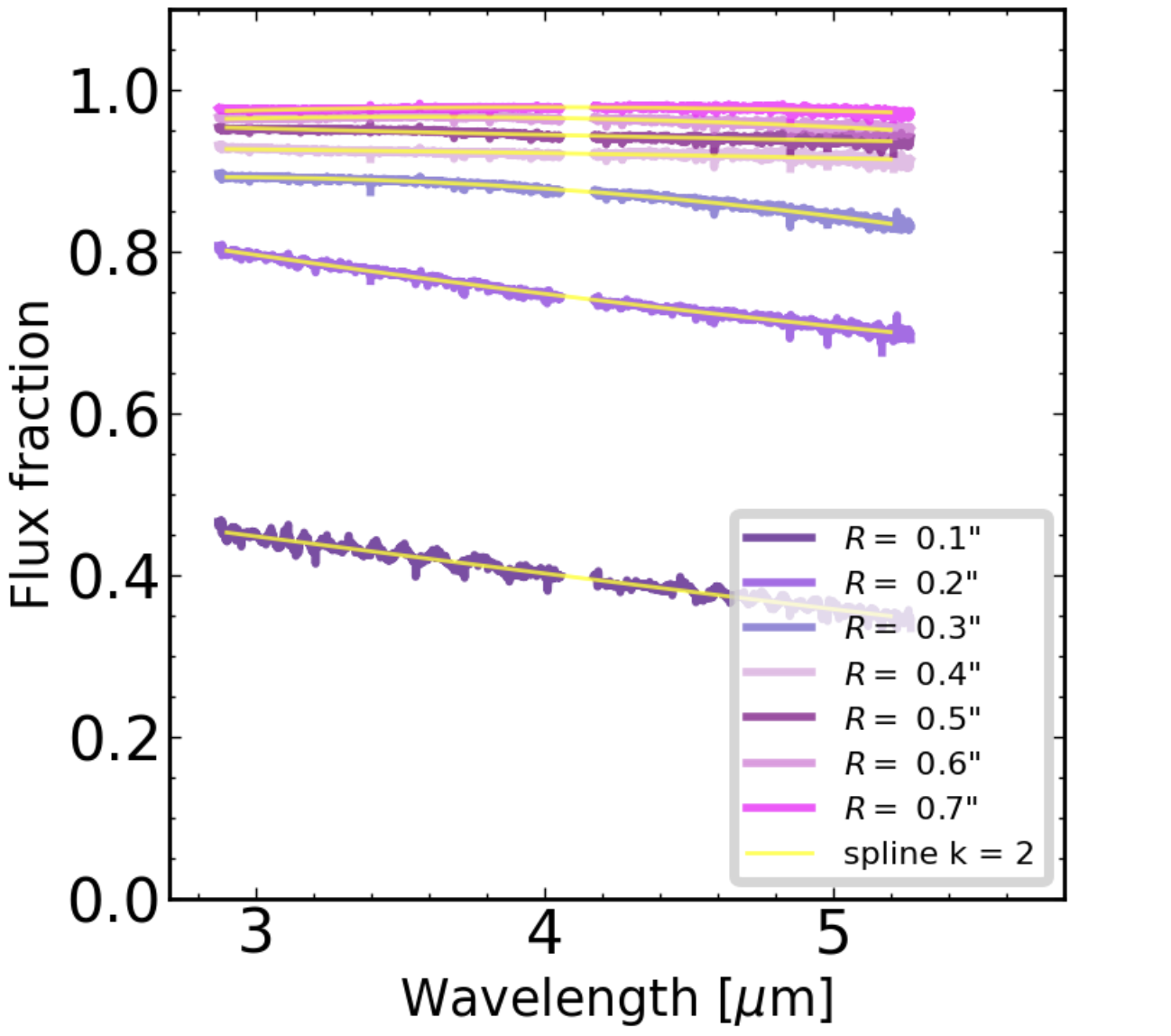}\\
\end{center}
\caption{\textit{Left panel}: flux fraction as a function of the aperture radius (NIRSpec/IFU, G395H/F290LP), i.e., the cumulative PSF. Note that the PSF width increases going from blue to red wavelengths. The differences are larger for small radii, as the PSF profile is steeper. The curves are based on observations of the G--dwarf GSPC P330-E used for a Cycle 1 calibration program (ID: 1538). \textit{Right panel}: Flux fractions for apertures with radii $R = 0.1\arcsec - 0.7\arcsec$ as a function of the observed wavelength (NIRSpec/IFU, G395H/F290LP). The second-order spline fitted to the data is shown in yellow. The flux fraction decreases with increasing wavelength, especially at small radii.}
\label{fig_psfProf}
\end{figure*}
\end{appendix}
\end{document}